\documentclass[aps,prl,reprint,preprintnumbers,showpacs,showkeys,superscriptaddress]{revtex4-2}
\usepackage{amsmath,amssymb,bm,mathrsfs,braket,color,ulem,mathrsfs}
\usepackage{graphicx}
\usepackage{srcltx}
\usepackage{color}
\definecolor{LinkColor}{RGB}{46,48,146}
\usepackage{hyperref}
\hypersetup{
colorlinks=true,
citecolor=LinkColor,
linkcolor=LinkColor,
urlcolor=LinkColor
}

\begin{document}
\title{Dynamical Orbital Angular Momentum Induced by Circularly Polarized Phonons}

\author{Dapeng Yao}
\affiliation{RIKEN Center for Emergent Matter Science (CEMS), 2-1 Hirosawa, Wako, Saitama, 351-0198, Japan}

\author{Dongwook Go}
\affiliation{Department of Physics, Korea University, Seoul 02841, Republic of Korea}
\affiliation{Peter Gr\"{u}nberg Institut and Institute for Advanced Simulation, Forschungszentrum J\"{u}lich and JARA, 52425 J\"{u}lich, Germany}
\affiliation{Institute of Physics, Johannes Gutenberg University Mainz, 55099 Mainz, Germany}

\author{Yuriy Mokrousov}
\affiliation{Peter Gr\"{u}nberg Institut and Institute for Advanced Simulation, Forschungszentrum J\"{u}lich and JARA, 52425 J\"{u}lich, Germany}
\affiliation{Institute of Physics, Johannes Gutenberg University Mainz, 55099 Mainz, Germany}

\author{Shuichi Murakami}
\affiliation{Department of Applied Physics, University of Tokyo, 7-3-1 Hongo, Bunkyo-ku, Tokyo 113-8656, Japan}
\affiliation{International Institute for Sustainability with Knotted Chiral Meta Matter (WPI-SKCM$^\text{2}$),
Hiroshima University, 1-3-1 Kagamiyama, Higashi-Hiroshima, Hiroshima 739-8526, Japan
}
\affiliation{RIKEN Center for Emergent Matter Science (CEMS), 2-1 Hirosawa, Wako, Saitama, 351-0198, Japan}

\begin{abstract}
We show that the orbital angular momentum (OAM) of electrons is dynamically induced by circularly polarized phonons. The induced OAM originates from the adiabatic evolution in which electrons acquire Berry phase formulated in terms of the Berry curvature encoded in phonon displacement space. By introducing a tight-binding model with $p$ orbitals on a honeycomb lattice, we show a microscopic picture that ionic rotations modulate orbital overlaps of electrons, and calculate the generated OAM, whose sign depends on phonon chirality. We then construct an effective model for valley phonons with different phonon pseudoangular momenta (PAM) and identity their distinct intervalley-scattering channels. Our model obeys the selection rule between phonons and electrons with the orbital degree of freedom. Extending this framework to $d$-orbital electrons, our model is applied to describe the induced OAM in monolayer transition metal dichalcogenides.
Our results reveal a direct orbital generation mechanism that emerges even in materials with weak spin-orbital coupling, opening a new promising way for orbitronics applications.
\end{abstract}
\maketitle

The discovery of circularly polarized phonons, which carry phonon angular momentum (PhAM), has triggered an intensive study on utilizing phonon-mediated angular momentum transport effects~\cite{Zhang2014,Zhang2015,Zhang2022,Tsunetsugu2023,Zhang2023,Zhang2025,S_Zhang2026,Zhu2018,ChenXT2019,Grissonnanche2020,Tauchert2022,Ueda2023,Ishito2023,Ohe2024,Zhang2025NC,Suzuki2025}.
Since PhAM shares the same axial-vector symmetry as magnetization, direct coupling between phonons and magnetism leads to novel magnetic responses via the couplings with electrons~\cite{Juraschek2017,Juraschek2019,Juraschek2022,Luo2023,ZhangCPL2023,Shabala2024}, spins~\cite{Bonini2023,Wu2023,Ren2024,Lujan2024,Wu2025,Che2025,Yang2025,Royo2025} and orbitals~\cite{Chaudhary2024,Mustafa2025}. In turn, such phonons with PhAM behave as a driving field, enabling the response of electron charge current~\cite{HamadaPRB2018,Yao2022,Yokoyama2025}, spin~\cite{Hamada2020,Fransson2023,Kim2023,Yao2024,YaoAPL2024,Yokoyama2024,Funato2024,Yao2025} and orbital magnetizations~\cite{Xiao2021,Ren2021,ZhangXW2023,Urazhdin2025}, magnon-phonon polarons~\cite{Go2019,Godejohann2020,Vaclavkova2021,Bao2023}, and thermal effect~\cite{HamadaPRL2018,Park2020,Ishizuka2025}.

Moreover, recent studies highlight the potential of orbitronics, with theoretical proposals~\cite{Bernevig2005,Tanaka2008,Go2018,Jo2018}, and experimental evidence~\cite{Choi2023,Lyalin2023} that shows strong electronic orbital responses even in non-relativistic regimes, overcoming many limitations of spin-based electronics. These include orbital currents driven by electric field~\cite{Go2020}, orbital-light interactions~\cite{Xu2025}, magnon-mediated orbital transport~\cite{Fishman2022,Go2024,An2015,DYao2025}, orbital pumping~\cite{Han2022,Hayashi2024,Go2025,Pezo2025}, and orbital torques on magnetization~\cite{DGo2020,Go2025}. Exploiting circularly polarized phonons to orbital generations provides a promising route toward novel functionalities in a hybridized scheme.

In this Letter, we show that ionic rotations can directly generate the electronic orbital angular momentum (OAM) originating from dynamically acquired Berry phase of electron orbitals~\cite{Berry1984}. We consider a simple microscopic picture that electronic overlaps with $p$ orbitals are dynamically modulated by ionic rotations, which is interpreted as an orbital electron-phonon coupling, leading to a nonzero electronic OAM. When the phonon displacement is small, the time-averaged OAM is formulated by the Berry curvature defined in the phonon displacement space. At the valley points, the phonons with finite momentum give rise to intervalley scatterings of electrons, which obeys the selection rule between phonons and electrons with orbital degree of freedom. We construct an effective model which is classified by phonon pseudoangular momenta (PAM), and identity their distinct intervalley-scattering channels.
Such methods can be also applied to the case of $d$-orbital electrons, by which we  describe the monolayer transition metal dichalcogenides (TMDs), and show the orbital generation in these materials.

\begin{figure}
\begin{center}
\includegraphics[width=\columnwidth]{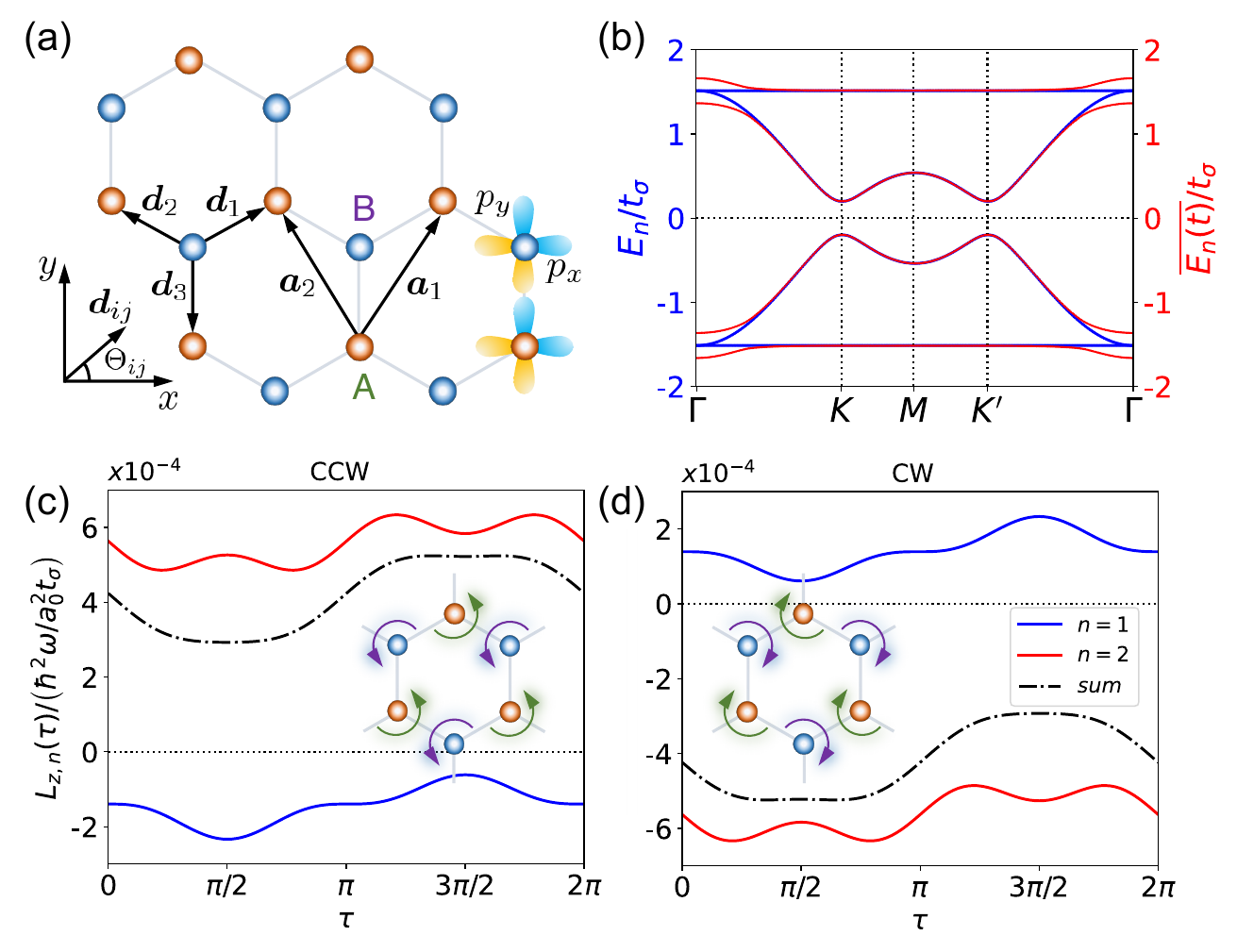}
\caption{Electronic states with phonon dynamics on 2D honeycomb lattice. (a) Schematic view of the 2D honeycomb lattice with $p_x$ and $p_y$ orbitals. (b) Electronic band structure without phonons blue lines) and that with the circularly polarized phonon at $\Gamma$ point (red lines) after taking time average. (c) and (d) Dynamical OAM induced by the optical-$\Gamma$ phonons. We show its dependence on the dimensionless time $\tau$ with the CCW phonon mode in the inset of (c) and CW phonon mode in the inset of (d). Red and blue solid lines represent the contributions from the first and second band shown in (b) below $E=0$, and the black dot-dashed line is their sum. Here we employ the parameters: $\Delta=0.2t_{\sigma}$ and $u_r=0.05a_0$.}
\label{fig1}
\end{center}
\end{figure}

\textit{Microscopic origin.---}
We consider a two-dimensional (2D) honeycomb lattice with $p_x$ and $p_y$ orbitals as shown in Fig.~\ref{fig1}(a). When circularly polarized phonons are switched on, the overlaps between $p_x$-$p_x$, $p_x$-$p_y$, and $p_y$-$p_y$ electronic orbitals are dynamically modulated by ionic rotations.
The electronic tight-binding (TB) Hamiltonian is given by
\begin{align}\label{Hamil}
\hat{H}_0=\sum_{\braket{ij}}\sum_{\alpha\beta}t_{ij}^{\alpha\beta}\hat{c}^{\dagger}_{i\alpha}\hat{c}_{j\beta}+\sum_{i}\sum_{\alpha}\Delta\xi_{i}\hat{c}^{\dagger}_{i\alpha}\hat{c}_{i\alpha},
\end{align}
where $\hat{c}^{\dagger}_{i\alpha}(\hat{c}_{j\beta})$ denotes the creation (annihilation) operator of the electron on the site $i(j)$ with $\alpha,\beta=x,y$ being the $p_x$ and $p_y$ orbitals. The first term represents the nearest-neighbor (NN) $\sigma$-type hopping, given by the Slater-Koster (SK) parameters
$t^{xx}_{ij}=\tilde{\eta}_{\sigma}\cos^2\Theta_{ij}/d_{ij}^2$,
$t^{xy}_{ij}=\tilde{\eta}_{\sigma}\cos\Theta_{ij}\sin\Theta_{ij}/d_{ij}^2$, and
$t^{yy}_{ij}=\tilde{\eta}_{\sigma}\sin^2\Theta_{ij}/d_{ij}^2$. Here $\tilde{\eta}_{\sigma}$ represents the parameter of the $\sigma$-type hopping~\cite{Harrison1994}, and $\bm d_{ij}$ is the vector from the site $i$ to site $j$ with $d_{ij}$ and $\Theta_{ij}$ denoting the length and angle of $\bm d_{ij}$ depicted in Fig.~\ref{fig1}(a). The second term describes a staggered on-site potential $\Delta$ with $\xi_{A(B)}=\pm1$ for the A(B) sublattices.

To provide a simple microscopic description, we first focus on the optical phonon modes at $\Gamma$ point, where the circularly polarized modes are superpositions of the degenerate modes.
In these optical modes, the atoms A and B rotate around their equilibrium positions with a phase difference of $\pi$.
They are classified into counterclockwise (CCW) and clockwise (CW) modes as shown in the inset of Figs.~\ref{fig1}(c) and~\ref{fig1}(d), respectively.
In this case, $d_{ij}$ and $\Theta_{ij}$ appearing in $t_{ij}^{\alpha\beta}$ change with time $t$, and these hopping parameters acquire time dependence. If the phonon displacements are small, this time dependence is written as $t_{ij}^{\alpha\beta}\rightarrow t_{ij}^{\alpha\beta}+\delta t_{ij}^{\alpha\beta}(t)$. Here the parameters are given by
$\delta t^{xx}_{ij}(t)=2t_{\sigma}\cos\Theta_{ij}u_x(t)/a_0-4t_{\sigma}\cos^2\Theta_{ij}\bm d_{ij}\cdot\bm u(t)/a_0^2$,
$\delta t^{yy}_{ij}(t)=2t_{\sigma}\sin\Theta_{ij}u_y(t)/a_0-4t_{\sigma}\sin^2\Theta_{ij}\bm d_{ij}\cdot\bm u(t)/a_0^2$, and
$\delta t^{xy}_{ij}(t)=t_{\sigma}\left[u_x(t)\sin\Theta_{ij}+u_y(t)\cos\Theta_{ij}\right]/a_0-4t_{\sigma}\cos\Theta_{ij}\sin\Theta_{ij}\bm d_{ij}\cdot\bm u(t)/a_0^2$, where $t_{\sigma}\equiv\tilde{\eta}_{\sigma}/a_0^2$ with $a_0$ being the equilibrium bond length, and $\bm u(t)=\bm u_B(t)-\bm u_A(t)$
denotes the relative phonon displacement~\cite{SM}.
In consequence, the electronic Hamiltonian modulated by ionic rotations acquires a dynamical term as
\begin{align}
\delta\hat{H}(t)=\sum_{\braket{ij}}\sum_{\alpha\beta}\delta t_{ij}^{\alpha\beta}(t)\hat{c}^{\dagger}_{i\alpha}\hat{c}_{j\beta},
\end{align}
which stands for the electron-phonon coupling. As an example, in the case of the CCW mode, the relative phonon displacement is given by $\bm u(t)=u_r(\cos\omega t,\sin\omega t)$, where $u_r=u_B-u_A$ is the relative amplitude of phonons with $u_A$ and $u_B$ being the rotating amplitude of the atoms A and B, and $\omega$ is the phonon frequency.
We notice that the modulated hopping parameter $\delta t_{ij}^{\alpha\beta}\sim(u_r/a_0)t_{ij}^{\alpha\beta}$ is of the first order in the relative amplitude.
Here we compare the band structures without phonons (blue lines) and those with phonons (red lines) after taking the time average in Fig.~\ref{fig1}(b), where two flat bands appear when phonons are absent due to the destructive interference of electronic waves on the honeycomb lattice with the $\sigma$-type hopping only~\cite{Wu2007,Wu2008,Liu2014}. A small energy shift between the bands with the red and blue lines comes from the energy transfer between phonons and electrons at the initial transient stage, and it will eventually approach a nonequilibrium steady state. Eventually, the electronic energy becomes a periodic function of time $t$: $E(t+T)=E(t)$ with the time period $T$.

\textit{Dynamical OAM with a geometric nature.---}
The overlap between electronic orbitals are periodically modulated by circularly polarized phonons. We assume that the phonon frequency is much smaller than the electronic band gap. By treating the atomic rotation as an adiabatic process, the electronic OAM is directly induced due to the geometric effect originating from the dynamically acquired Berry phase~\cite{Berry1984,Trifunovic2019}, and the time-dependent OAM for the $i(=x,y,z)$ component of the $n$th band at time $t$ is given by~\cite{SM}
\begin{align}\label{L_znt}
L_{i,n}(t)=\int_{\bm k}\sum_{m(\neq n)}\left\{\frac{\hbar\hat{L}_{i,nm}(t)A_{mn}(t)}{E_n(t)-E_m(t)}+\text{c.c}\right\},
\end{align}
where $\int_{\bm k}\equiv\int_{\text{BZ}}\frac{d\bm k}{(2\pi)^2}$ is the integration over the 2D Brillouin zone (BZ) of electrons, $\hat{L}_{i,nm}(t)=\bra{\psi_n(t)}\hat{L}_{i}\ket{\psi_m(t)}$ is the instantaneous matrix element with $\hat{L}_{i}$ being the $i$-component of the OAM operator, $A_{mn}=i\braket{\psi_m(t)|\partial_t\psi_n(t)}$ is the instantaneous Berry connection with the eigenstate $\psi_n(t)$ of the $n$th band at time $t$, and $E_n(t)$ is the eigenvalue of the eigenstate $\psi_n(t)$.
In our $p_x$-$p_y$ model, only the out-of-plane component of OAM is nonzero, and its operator is given by $\hat{L}_z=\hbar s_0\otimes\sigma_y$ in the basis: $(\ket{A,p_x},\ket{A,p_y},\ket{B,p_x},\ket{B,p_y})$, with the identity matrix $s_0$ and Pauli matrix $\sigma_y$.
We show the dynamical OAM induced by the phonons with CCW and CW modes in Figs.~\ref{fig1}(c) and~\ref{fig1}(d) during a time period by replacing the time $t$ by a dimensionless time $\tau=\omega t$ for simplicity, and we note that their time averages are nonzero.
By taking the phonon energy $\hbar\omega=0.02$eV~\cite{Li2019,Bae2022}, the hopping parameter $t_{\sigma}=1$eV, and the lattice constant $a_0=2$\AA, we estimate the time-averaged OAM to be $10^{-5}\mu_B$ per unit cell.

Here, we can formulate the time-averaged OAM. For simplicity, for the moment we assume that the crystal has only one species of atoms with the in-plane displacement $\bm u=(u_x,u_y)$. We introduce an orbital Zeeman field $B_{i}$ as a conjugate field to the OAM, i.e., the OAM in the absence of $B_{i}$ is given by $\hat{L}_{i}=\frac{\hbar}{\mu_B}\partial_{B_{i}}\hat{H}|_{\bm B=0}$ with Bohr magneton $\mu_B$ and the total Hamiltonian $\hat{H}$. Then the Berry connection can be replaced by $A_{mn}=i\braket{\psi_m|\partial_{\bm u}\psi_n}\cdot\dot{\bm u}$. When the displacement $\bm u$ is small compared with the lattice constant, the time-averaged OAM is expanded near $\bm u=0$ and formulated as~\cite{SM}
\begin{align}\label{Lave}
\bar{L}_{i,n}=\frac{\hbar^2}{2\mu_B}J^{\text{ph}}_{z}\int_{\bm k}\partial_{B_{i}}\Omega^{(n)}_{u_xu_y}\Big|_{\bm u=0,\bm B=0},
\end{align}
where $J^{\text{ph}}_{z}=\frac{1}{T}\int_0^Tdt(\bm u\times\dot{\bm u})_z$ denotes the PhAM divided by the atomic mass, and $\Omega^{(n)}_{u_xu_y}\equiv\partial_{u_x}A_{u_y}^{(n)}-\partial_{u_y}A_{u_x}^{(n)}$ represents the Berry curvature with $A^{(n)}_{u_i}=i\braket{\psi_n|\partial_{u_i}\psi_n}$ in terms of the displacement. This derivation is similar to that for the spin angular momentum~\cite{Yao2025}.

\begin{figure}
\begin{center}
\includegraphics[width=\columnwidth]{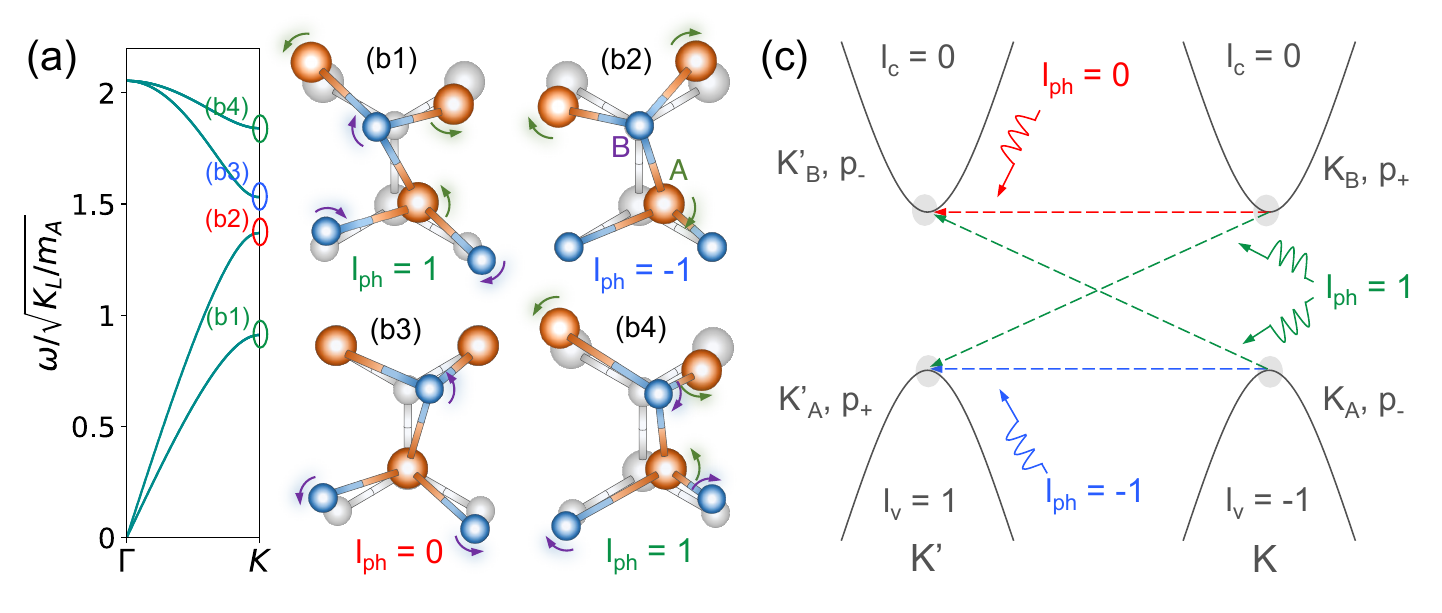}
\caption{Valley phonons and selection rule. (a) Phonon dispersion between the high-symmetry points $\Gamma$ and $K$ in the phonon BZ. (b) Phonon eigenmodes at $K$ point with PAM. Here we set the location of the atom A marked in (b2) as the center of $C_3$ rotation. We assume that the springs only exist between the NN atoms and characterized by two stiffness constants $K_T$ and $K_L$, which describe the stiffness against the deformations perpendicular and parallel to the spring, respectively. Here we take $K_T=K_L/4$, and the masses of atoms A and B satisfies $m_B=0.8m_A$~\cite{SM}. (c) Phonons with different PAM lead to intervalley scattering of electrons, which satisfy the selection rules between phonons and electrons with orbital degree of freedom.
}
\label{fig2}
\end{center}
\end{figure}

\textit{Effect of valley phonons and selection rule.---}
Next, we consider the phonons at the $K$ point.
Here a phonon displacement polarization vector $\bm\epsilon_{\bf k}$ is an eigenstate of $C_3$ rotation symmetry: $\mathcal{\hat{R}}_3\bm\epsilon_{\bf k}=e^{-i\frac{2\pi}{3}l_{\text{ph},\bf k}}\bm\epsilon_{\bf k}$, where $\mathcal{\hat{R}}_3$ denotes $C_3$ rotation operator acting on the phonon mode,  $\bf k$ is limited to $C_3$ invariant momenta, and $l_{\text{ph},\bf k}$ represents the phonon PAM~\cite{Zhang2015,Zhang2022}. 
We show the schematic pictures of the phonon modes with PAM at $K$ point in Figs.~\ref{fig2}(b1)-(b4), 
where all the phonon modes are circularly polarized.
The center of $C_3$ rotation is the location of the atom A. These phonon modes directly modulate the next-nearest-neighbor (NNN) electronic hoppings since the same species of atoms rotate with a phase difference $e^{\pm i2\pi/3}$.
One can calculate the electronic TB Hamiltonian after enlarging the unit cell by three times, and the electronic Bloch states at the $K$ and $K'$ points are folded onto the $\Gamma$ point~\cite{Ren2015,SM}.
Instead, here we construct an effective model of electrons to describe the effect of phonons at $K(K')$ point.
At the valley points of electrons with $p$ orbitals, it is convenient to change the basis from $p_x/p_y$ to $p_{\pm}=p_x\pm ip_y$, since the Bloch states with $p_{\pm}$ are $C_3$ eigenstates~\cite{SM}.

The effective Hamiltonian of electrons near the $\Gamma$ point with phonons at $K$ point is classified by the phonon PAM as shown in Figs~\ref{fig2}(b1)-(b4). Here we choose the basis as $(\ket{K_A,p_-},\ket{K_B,p_+},\ket{K'_A,p_+},\ket{K'_B,p_-})$, and the $C_3$ rotation operation is represented by $D(C_3)=\text{diag}(e^{i\frac{2\pi}{3}},1,e^{-i\frac{2\pi}{3}},1)$ with the location of atom A being the rotation center.
First, in the case of phonons with $l_{\text{ph}}=1$ shown in Figs.~\ref{fig2}(b1) and~\ref{fig2}(b4), the atoms A and B rotate in opposite directions. The effective Hamiltonian is given by
\begin{align}\label{eff_H_l=1}
\mathcal{H}_{\text{eff}}^{(l_{\text{ph}}=1)}=
\begin{bmatrix}
\Delta & -\hbar v_F\xi & 0 & \lambda\rho \\
-\hbar v_F\xi^{\dagger} & -\Delta & \lambda\rho & 0 \\
0 & \lambda\rho^{\dagger} & \Delta & \hbar v_F\xi^{\dagger} \\
\lambda\rho^{\dagger} & 0 & \hbar v_F\xi & -\Delta
\end{bmatrix},
\end{align}
where $\xi=q_x+iq_y$ with $q_a$ $(a=x,y)$ being a small wavenumber near the $\Gamma$ point, and $v_F=3a_0t_{\sigma}/2\hbar$ denotes the Fermi velocity. The phonon with $l_{\text{ph}}=1$ leads to an intervalley-scattering term: $\rho=-u_y+v_y+i(u_x+v_x)$ with the coefficient $\lambda=-3t_{\sigma}/a_0$, where $(u_x,u_y)$ and $(v_x,v_y)$ denote the displacements of atoms A and B, respectively.
Next, in the cases of phonons with $l_{\text{ph}}=-1,0$ shown in Figs.~\ref{fig2}(b2) and~\ref{fig2}(b3), either the atom A or B makes a circular motion. The effective Hamiltonian with $l_{\text{ph}}=-1$ reads
\begin{align}\label{eff_H_l=-1}
\mathcal{H}_{\text{eff}}^{(l_{\text{ph}}=-1)}=
\begin{bmatrix}
\Delta & -\hbar v_F\xi & \mu\chi_A & 0 \\
-\hbar v_F\xi^{\dagger} & -\Delta & 0 & 0 \\
\mu\chi_A^{\dagger} & 0 & \Delta & \hbar v_F\xi^{\dagger} \\
0 & 0 & \hbar v_F\xi & -\Delta
\end{bmatrix},
\end{align}
and that with $l_{\text{ph}}=0$ takes the form as
\begin{align}\label{eff_H_l=0}
\mathcal{H}_{\text{eff}}^{(l_{\text{ph}}=0)}=
\begin{bmatrix}
\Delta & -\hbar v_F\xi & 0 & 0 \\
-\hbar v_F\xi^{\dagger} & -\Delta & 0 & \mu\chi_B \\
0 & 0 & \Delta & \hbar v_F\xi^{\dagger} \\
0 & \mu\chi_B^{\dagger} & \hbar v_F\xi & -\Delta
\end{bmatrix},
\end{align}
where $\mu=-6t'_{\sigma}/a_0$ with $t'_{\sigma}$ being the $\sigma$-type hopping parameter between the NNN atoms~\cite{SM}. The intervalley-scattering terms are given by $\chi_A=-u_y-iu_x$ and $\chi_B=v_y+iv_x$, which come from the direct modulation of NNN electronic hoppings between the same species of atoms.

\begin{figure}
\begin{center}
\includegraphics[width=\columnwidth]{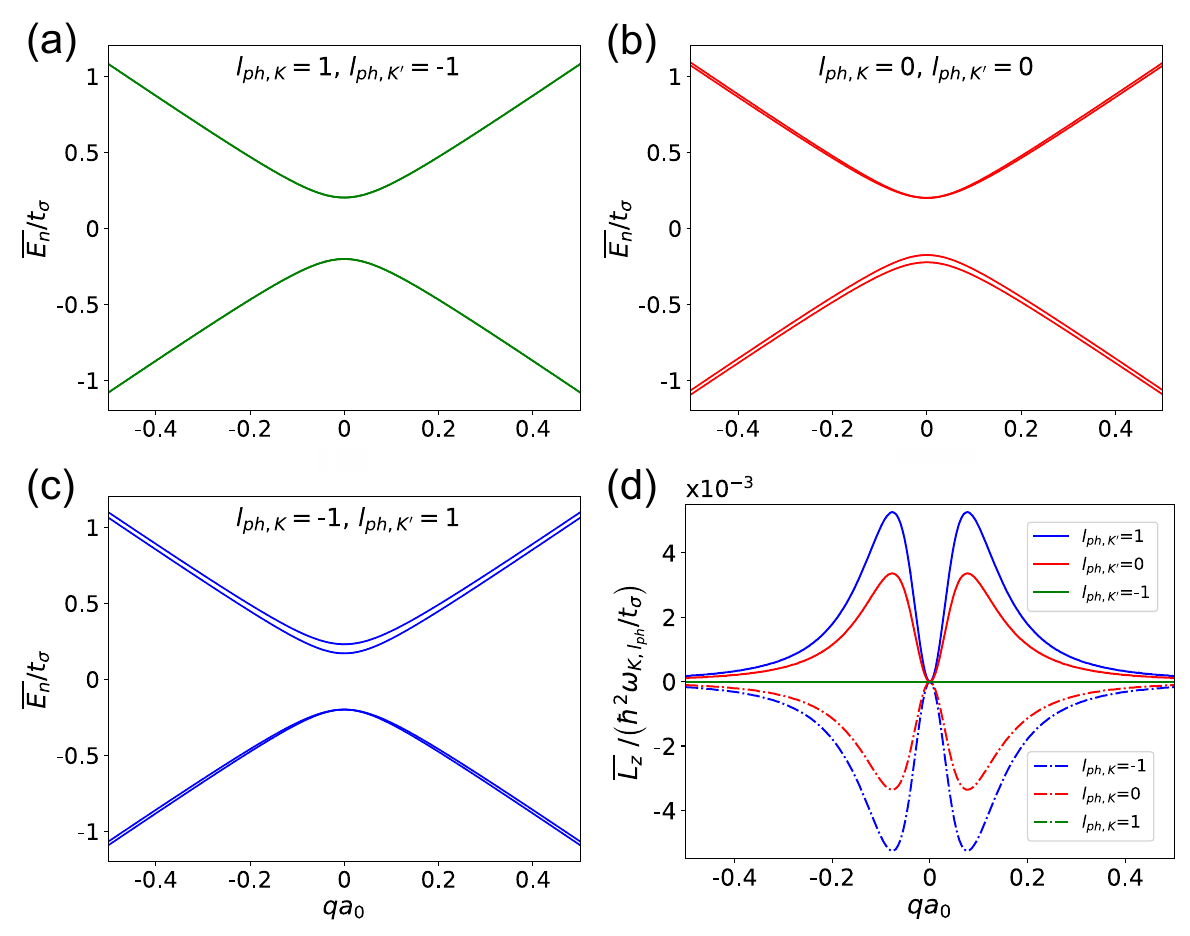}
\caption{Band structure and OAM calculated from the effective Hamiltonian with $p$ orbitals. Band structure after taking time average calculated from the effective Hamiltonian plotted by (a) green lines with $l_{\text{ph},K}=1$ and $l_{\text{ph},K'}=-1$, (b) red lines with $l_{\text{ph},K}=0$ and $l_{\text{ph},K'}=0$, and (c) blue lines with $l_{\text{ph},K}=-1$ and $l_{\text{ph},K'}=1$. (d) Time-averaged OAM induced by phonons with different PAM. Here the horizontal axis is the wavenumber near the $\Gamma$ point. The parameters are employed as $\Delta=0.2t_{\sigma}$, $t'_{\sigma}=0.1t_{\sigma}$, $u_A=3u_B=0.05a_0$ with $a_0=1$.
}
\label{fig3}
\end{center}
\end{figure}

The effective Hamiltonian distinguished by the PAM can be understood from the selection rule for phonons and electrons with the orbital degree of freedom. As shown in Fig.~\ref{fig2}(c), the pseudoangular momenta of electrons in conduction and valence bands are obtained from the $C_3$ rotation representation: $l_c(K)=l_c(K')=0$ and $l_v(K/K')=\mp 1$. The intervalley-scattering terms $\rho$, $\chi_{A}$ and $\chi_{B}$ appearing in the effective Hamiltonian from the phonons at $K$ point with phonon PAM $l_{\text{ph}}=1,0,-1$ satisfy the selection rule: $l_{v(c)}(K')-l_{v(c)}(K)=l_{\text{ph}}(K)~(\text{mod}~3)$, which comes from the momentum conservation $\bm k_{K}+\bm k_{K}=\bm k_{K'}$.

We first show the time-averaged bands calculated from the effective Hamiltonians with different PAM in Figs.~\ref{fig3}(a)-(c). In the case of $l_{\text{ph},K}=1$, each band is doubly degenerate as shown in Fig.~\ref{fig3}(a), which consists of the electronic Bloch states at $K$ and $K'$ points. On the other hand, the band splitting occurs for $l_{\text{ph},K}=0,-1$ as shown in Figs.~\ref{fig3}(b) and~\ref{fig3}(c).
We next consider the electronic OAM operator $\hat{L}_z$, whose eigen equation at the $K$ point reads $\hat{L}_z\ket{K,p_{\pm}}=\pm\hbar\ket{K,p_{\pm}}$.
The electronic Bloch state at the $K'$ point is given by $\ket{K',p_{\pm}}=\hat{\Theta}\ket{K,p_{\mp}}$ via the time-reversal operation $\hat{\Theta}$.
Since the OAM operator satisfies $\hat{\Theta}\hat{L}_z\hat{\Theta}^{-1}=-\hat{L}_z$, the eigen equation of $\hat{L}_z$ at the $K'$ point yields $\hat{L}_z\ket{K',p_{\pm}}=\pm\hbar\ket{K',p_{\pm}}$.
In the same basis with the effective Hamiltonian, the OAM operator is given by $\hat{L}_z=-\hbar s_z\otimes\sigma_z$.
We then show the time-averaged OAM with different PAM evaluated by Eq.~(\ref{L_znt}) in Fig.~\ref{fig2}(d) as a function of the isotropic wavenumber $q=\sqrt{q_x^2+q_y^2}$.
The OAM for $l_{\text{ph},K}=-1,0$ show peaks on both sides of $q=0$ while that for $l_{\text{ph},K}=1$ is always zero~\cite{SM}.
Furthermore, the phonons at $K'$ point are mutually related to those at $K$ point by TRS, and the PAM at $K$ and $K'$ points are always opposite~\cite{Zhang2015}. We also show the OAM induced by phonons at $K'$ point, which have an opposite sign from those at $K$ point. It means that phonon chirality is reflected in the electronic states.

\begin{figure}
\begin{center}
\includegraphics[width=\columnwidth]{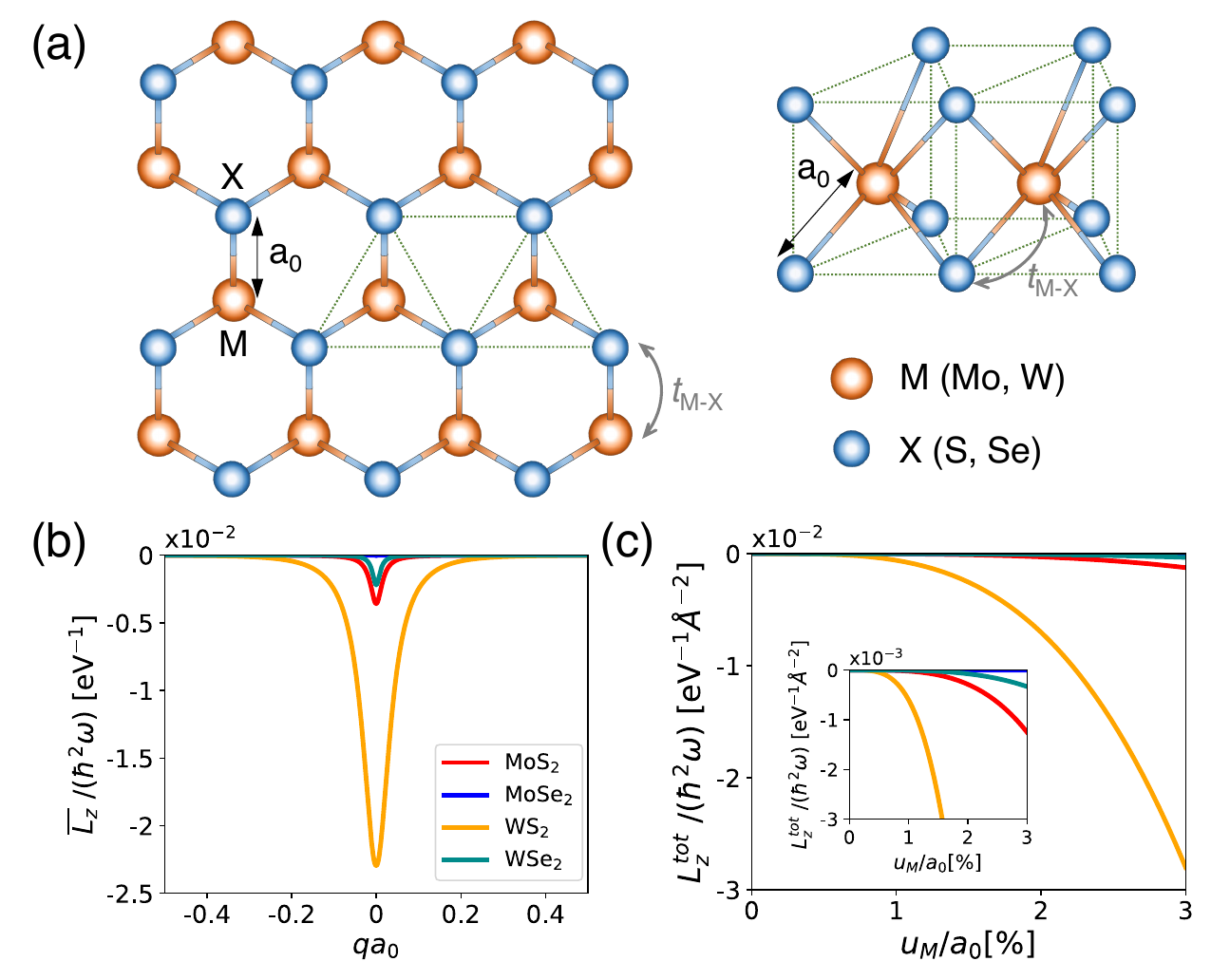}
\caption{Electronic OAM induced by the phonon at $K$ point with $l_{\text{ph}}=1$ in monolayer TMDs MoS$_2$, MoSe$_2$, WS$_2$, and WSe$_2$. (a) Schematic picture of a TMD crystal structure with $t_{\text{M-X}}$ and $a_0$ being the hopping parameter and the distance between the atoms M and X. The left side is the lattice from the top view. (b) OAM as a function of wavenumber $q$ near the $\Gamma$ point with $u_{\text{M}}/a_0=1\%$. (c) Total OAM as a function of $u_{\text{M}}/a_0$, which represent the size that the atom M displaced from the equilibrium.
}
\label{fig4}
\end{center}
\end{figure}

\textit{Application to 2D monolayer TMDs.---}
We generalize our effective Hamiltonian to the 2D monolayer TMDs MX$_2$ with representative examples being $\text{M}=\text{W},~\text{Mo}$ and $\text{X}=\text{Se},~\text{S}$. Here the monolayer unit is characterized by the honeycomb lattice composed of the atoms M and X as shown in Fig.~\ref{fig4}(a). The electronic Bloch states at $K(K')$ point consists of hybrids of $p$ orbitals of $\text{X}$ and $d$ orbitals of $\text{M}$~\cite{Liu2013,Fang2015}.
Therefore, we can construct an effective Hamiltonian for $d$ orbitals from symmetry analysis.
At the $K$ and $K'$ points, the conduction band mostly consists of the $d_{z^2}$ orbital while the valence band is mainly composed of the $d_{x^2-y^2}+id_{{xy}}$~($d_{x^2-y^2}-id_{{xy}})$ orbitals at the $K(K')$ point~\cite{Fang2015}.
Let $d_0\equiv d_{z^2}$ and $d_{\pm2}\equiv d_{x^2-y^2}\pm id_{{xy}}$, where the subscripts $0,\pm2$ represent the magnetic quantum numbers of these $d$ orbitals.
Now we introduce the phonons at $K$ point with $l_{\text{ph}}=1$. We choose the basis as $(\ket{K,d_{2}},\ket{K,d_0},\ket{K',d_{-2}},\ket{K',d_0})$, and the center of $C_3$ rotation located at the atom M. The representation matrix of the $C_3$ rotation is $D(C_3)=\text{diag}(e^{i\frac{2\pi}{3}},1,e^{-i\frac{2\pi}{3}},1)$, which is as same as that for the basis of the effective Hamiltonian with $p$ orbitals. Thus, the effective Hamiltonian for MX$_2$ with phonons at $K$ point takes the form as
\begin{align}\label{HMX2}
\mathcal{H}_{\text{MX$_2$}}^{(l_{\text{ph}}=1)}=\begin{bmatrix}
\epsilon_2 & -\tilde{v}\xi & 0 & \tilde{\lambda}\rho \\
-\tilde{v}\xi^{\dagger} & \epsilon_0 & \tilde{\lambda}\rho & 0 \\
0 & \tilde{\lambda}\rho^{\dagger} & \epsilon_2 & \tilde{v}\xi^{\dagger} \\
\tilde{\lambda}\rho^{\dagger} & 0 & \tilde{v}\xi & \epsilon_0
\end{bmatrix},
\end{align}
where $\epsilon_0$ and $\epsilon_2$ denote the energy of $d_0$ and $d_{2}$ orbitals at the $K$ point, respectively.
Here $\tilde{v}=3a_0t_{\text{M-X}}/2$ with $t_{\text{M-X}}$ denoting the hopping parameter between the $d_{xy}$ orbital of the atom M and the $p_z$ orbital of the atom X, and $\tilde{\lambda}=-3t_{\text{M-X}}/a_0$ with $a_0$ being the bond length between the atoms X and M shown in Fig.~\ref{fig4}(a) (see End Matter for details).
We then introduce the time-dependent displacements of the lowest phonon mode at $K$ point as $(u_x,u_y)=u_{\text{M}}(\cos{\tau},\sin{\tau})$ of atom M and $(v_x,v_y)=u_{\text{X}}(-\cos{\tau},\sin{\tau})$ of atom X with their rotating amplitude $u_{\text{M}}$ and $u_{\text{X}}$, and we then have $\tilde{\lambda}\rho=-i\tilde{\lambda}_ue^{i\tau}$ with $\tilde{\lambda}_u=\tilde{\lambda}(u_{\text{X}}-u_{\text{M}})$.

Under the basis for $d$ orbitals, the OAM operator is given by $\hat{L}_z=\text{diag}(2\hbar,0,-2\hbar,0)$. By using Eq.~(\ref{L_znt}), the OAM induced by the phonon at $K$ point with PAM $l_{\text{ph}}=1$ with phonon frequency $\omega$ as a function of $q$ is given by
\begin{align}
\overline{L}_z(q)=\frac{-\tilde{\lambda}_u^4\hbar^2\omega}{(\tilde{v}^2q^2+\Delta^2+\tilde{\lambda}_u^2)^{\frac{3}{2}}(\tilde{v}^2q^2+\tilde{\lambda}_u^2)}
\end{align}
with $\Delta=\frac{\epsilon_2-\epsilon_0}{2}$. 
We show the results for the typical 2D monolayer TMDs in Fig.~\ref{fig4}(b) by taking $u_{\text{M}}/a_0=1\%$.
Meanwhile, after integrating the contributions over the wavenumbers $q_x$ and $q_y$ by extending the region of integration to infinity, the total OAM can be obtained as $L_z^{\text{tot}}=-(2\pi\tilde{\lambda}_u^4\hbar^2\omega/|\Delta|^3\tilde{v}^2)\text{ln}|2\Delta/\tilde{\lambda}_u|$.
Our result shows the inverse cubic dependence on the energy gap, which comes from the nature of derivative of Berry curvature in Eq.~(\ref{Lave})~\cite{SM}. This gap dependence is seen in the intra-atomic OAM, whereas inter-atomic OAM shows an inverse quadratic gap dependence~\cite{Ren2021,Xiao2021}.
Therefore, the induced OAM in our work is smaller than that from inter-atomic mechanism. Nevertheless, our intra-atomic contribution becomes larger for materials with small gaps because of the $\Delta^{-3}$ dependence on the band gap.

We show the results for the TMDs as a function of $u_{\text{M}}/a_0$ in Fig.~\ref{fig4}(c).
Based on the experimental measurements~\cite{Li2019,Bae2022}, we assume that the phonon energy is $0.02$eV. Combining with our calculation, we find that the induced OAM converted to magnetic moment can reach 10$^{-4}\mu_{\text{B}}$ per unit cell, which is experimentally observable.
In the End Matter, we compare the OAM and magnetic moments of electrons and phonons in TMD materials. Despite the larger PhAM, the corresponding magnetic moments are comparable because ions have a much smaller gyromagnetic ratio than electrons.

\textit{Conclusion.---}
In this Letter, we show the dynamical OAM of electrons induced by circularly polarized phonons, which originates from the dynamically acquired Berry phase of electronic states modulated by ionic rotations.
We clarify the microscopic origin and establish a simple effective model to describe the response of electrons to phonons with orbital degree of freedom. Here the phonon-induced intervalley scattering of electronic orbitals obeys the selection rule, and classified by phonon PAM.
Our approach is generally applied to 2D monolayer materials, where the induced OAM can be easily estimated. The electronic OAM depends on phonon chirality, which enables engineering of orbital generation in orbitronics.

The previous study calculates the OAM for the phonons which are adiabatically switched on, by the time-dependent perturbation~\cite{Urazhdin2025}. We note that this approach is suited for few-level or molecular systems. In contrast, our method can study both crystalline and molecular systems. We formulate the OAM due to the geometric-phase accumulation of electrons during the phonon dynamics, and therefore, our method can calculate the time-dependence of the OAM over the phonon period. Furthermore, our method allows us to express its time average in a concise formula in terms of the Berry curvature involving phonon displacement space.

In nonmagnetic systems, the PhAM at time-reversal invariant momenta are opposite, resulting in a zero OAM when the two modes are equally populated~\cite{Zhang2015}. A finite net OAM can be obtained by phonon pumping with terahertz pulses~\cite{Nova2017}, which leads to an asymmetric population of phonons between CCW and CW modes.
Here the phonon amplitude assumed as 1\% of the bond length is unlikely to be sustained in steady states. Thus, our estimate based on steady states should be understood as a transient response under phonon pumping within initial few oscillatory cycles.
In addition, the coupling between phonons and electronic spin angular momentum requires spin-orbital coupling (SOC)~\cite{Hamada2020,Ohe2024,Yao2025}, while that between phonons and electronic OAM does not, and therefore the phonon-induced OAM appears even in the materials with weak SOC, such as a light metal titanium.
A practical detection scheme is to convert the induced OAM into an electrical signal via the inverse orbital Hall effect~\cite{Wang2023}. In a Hall-bar device with a sizable orbital Hall response, locally driven circular polarized phonons by terahertz pumping, generate a nonequilibrium orbital current. Because the induced OAM is time-reversal odd, reversing the phonon chirality flips the orbital current and the Hall voltage accordingly.

\textit{Note added.---}
We have become aware of a related work formulated the orbital accumulation induced by surface acoustical waves within longer wavelength limit~\cite{SatoOAM2025}. Our work focus on the circularly polarized phonons with a wavelength of the order of a lattice constant. By gradually changing the wavelength scale, we expect a crossover between the physics in our paper and Ref.~\cite{SatoOAM2025}.

\begin{acknowledgments}
\textit{Acknowledgments.---}
D.Y. acknowledges fruitful discussion with Shogo Yamashita during visiting Forschungszentrum J\"{u}lich.
D.G. acknowledges fruitful discussion with Junho Suh.
D.Y. was supported by Japan Society for the Promotion of Science (JSPS) KAKENHI Grant No.~JP23KJ0926, No.~JP25K23366, and RIKEN Special Postdoctoral Researchers Program.
Y.M. and D.G. acknowledge support by the Deutsche Forschungsgemeinschaft (DFG) 
in the framework of TRR 288$-$422213477 (Project B06), and by the EIC Pathfinder 
OPEN grant 101129641 ``OBELIX''.
D.G. was also supported by a Korea University Grant (K2528611).
S.M. was supported by JSPS KAKENHI Grant, No.~JP22H00108, No.~JP24H02231, and also by MEXT Initiative to Establish Next-generation Novel Integrated Circuits Centers (X-NICS) Grant No.~JPJ011438.
\end{acknowledgments}


%

\onecolumngrid \vspace{0.5cm}
 \begin{center}{\large\textbf{End Matter}}\end{center} \vspace{0.1cm}
\twocolumngrid \noindent

\textit{Parameters in effective model of TMDs.---}
The energy bands at $K/K'$ are composed of various orbitals such as $d_{z^2}$, $d_{x^2-y^2}$, and $d_{xy}$ orbitals of the atom M and $p_x$, $p_y$, and $p_z$ orbitals of atom X.
In our effective Hamiltonian for the 2D monolayer TMDs MX$_2$ in Eq.~(\ref{HMX2}), we focus on the two bands near the Fermi energy, which are the $d_0$ and $d_{\pm2}$ orbitals of atom M.
Here the electron hoppings between the NNN M atoms comes from two hopping processes: the direct NNN hopping and the NNN hopping by two NN hopping processes via the X atom, whose hopping parameters involve $t_{\text{M-M}}$ and $t_{\text{M-X}}$.
The latter type with the parameter $t_{\text{M-X}}$ [see Fig.~\ref{fig4}(a)] dominates the NNN hopping between the two NNN M atoms~\cite{Fang2015}.
In our model, the coefficient of the intervalley-scattering term $\tilde{\lambda}$ for the 2D monolayer MX$_2$ is given by $\tilde{\lambda}=-3t_{\text{M-X}}/a_0$, which has the same form with the term in the effective Hamiltonian with $p$ orbitals in Eq.~(\ref{eff_H_l=1}).
The simplest two-band model near the Fermi energy at the $K$ or $K'$ point is given by~\cite{Fang2015}
\begin{align}\label{H2K}
\mathcal{H}_{K/K'}(\bm q)=\frac{f_0}{2}\left(1+\sigma_z\right)+f_1a_0(\pm q_x\sigma_x+q_y\sigma_y),
\end{align}
with respect to the basis $(\ket{K/K',d_0},\ket{K/K',d_{\pm2}})$, where $f_0$, $f_1$, and $a_0$ are given in Table~\ref{tab:hopping} for various 2D monolayer materials~\cite{Fang2015}. It agrees with our effective Hamiltonian in Eq.~(\ref{HMX2}) without phonons.
Comparing our effective Hamiltonian in Eq.~(\ref{HMX2}) with the two-band Hamiltonian in Eq.~(\ref{H2K}), we find the values of the parameters: $\Delta=-f_0/2$, $\tilde{v}=-f_1a_0$, and $\tilde{\lambda}_u=-2\tilde{v}(u_{\text{X}}-u_{\text{M}})/a^2_0=2f_1(u_{\text{X}}-u_{\text{M}})/a_0$ for our effective Hamiltonian.
Here we assume that the phonon displacements are determined by the masses of atoms M and X as $u_{\text{X}}/u_{\text{M}}=m_{\text{M}}/m_{\text{X}}$. Here we also list these parameter values in Table~\ref{tab:hopping} with $u_{\text{M}}/a_0=1\%$.

\begin{table}[]
	\centering
	\caption{Parameters for various 2D monolayer TMDs MX$_2$. The upper three lines represent the lattice constant $a_0$ and the parameters $f_0$ and $f_1$~\cite{Fang2015}. By comparing our model with the model in Ref.~\cite{Fang2015}, we find the parameters in the lower three lines: $\Delta$, $\tilde{v}$, and $\tilde{\lambda}_u$ for our effective Hamiltonian in Eq.~(\ref{HMX2}). We assume that the displacement of atom M is 1\% of the bond length $a_0$.}
		\label{tab:hopping}
    \begin{tabular}{ccccc} \hline \hline
      & MoS$_2$ & MoSe$_2$ & WS$_2$ & WSe$_2$ \\ \hline    
    $a_0$(\r{A}) &  $2.41$ & $2.52$ & $2.42$ & $2.55$  \\
    $f_0$(eV) & $1.674$ & $1.442$ & $1.813$ & $1.546$ \\ 
    $f_1$(eV) & $1.152$ & $0.956$ & $1.407$ & $1.189$  \\ 
    \hline  
    $\Delta$(eV) &  $-0.837$ & $-0.721$ & $-0.907$ & $-0.773$ \\
    $\tilde{v}$(eV$\cdot$\r{A}) & $-2.776$ & $-2.409$ & $-3.405$ & $-3.032$ \\  
    $\tilde{\lambda}_u$(eV) & $0.046$ & $0.004$ & $0.133$ & $0.032$ \\
    \hline \hline     
      \end{tabular}
\end{table}

\textit{Phonon orbital magnetic moment of TMDs.---}
To compare the orbital magnetization between electrons and phonons, here we calculate the orbital magnetic momentum from the motions of ions. The time-dependent displacements of the lowest phonon modes at $K$ point of atoms M and X are $\bm u(t)=u_{\text{M}}(\cos{\omega t},\sin{\omega t})$ and $\bm v(t)=u_{\text{X}}(-\cos{\omega t},\sin{\omega t})$, respectively. The OAM of the two atoms are then given by $\bm L^{\text{M}}=m_{\text{M}}\bm u\times\dot{\bm u}=\omega m_{\text{M}}u^2_{\text{M}}\bm e_z$ and $\bm L^{\text{X}}=m_{\text{X}}\bm v\times\dot{\bm v}=-\omega m_{\text{X}}u^2_{\text{X}}\bm e_z$.
The magnetic moment of ion I($=\text{M,X}$) is given by
\begin{align}
\mu_{\text{I}}=\frac{eZ^*_{\text{I}}}{2m_{\text{I}}}L_z^{\text{I}},
\end{align}
where $Z^*_{\text{I}}$ represents the out-of-plane Born effective charges of the monolayer TMDs listed in Table.~\ref{tab:magnetic_moment}. Thus, the PhAM and phonon magnetic moment per unit cell are given by $\mu_{\text{ph}}=\mu_{\text{M}}+2\mu_{\text{X}}$ and $L_{\text{ph}}=L_{\text{M}}+2L_{\text{X}}$, respectively.

We show the results of the PhAM and electron OAM, and theirs converted to magnetic moments in Table~\ref{tab:magnetic_moment}. We find that even though the PhAM are much larger than electron OAM, their magnetic moments are comparable because the gyromagnetic ratios of ions are much smaller than that of electrons. The induced electron OAM proposed in this letter can be in the order of 10$^{-4}\mu_{\text{B}}$ per unit cell in WS$_2$. Even though experimentally separating the induced electron OAM and PhAM is difficult, the contribution only from ions is ignorable.

\begin{table}[]
	\centering
	\caption{Born effective charges and calculated OAM for 2D TMDs. The upper two lines represent the Born effective charges of monolayer TMDs~\cite{Sohier2016}. The lower two lines are the phonon orbital magnetic moment $\mu_{\text{ph}}$ and electron orbital magnetic moment $\mu_{\text{el}}$ per unit cell in the unit of Bohr magneton $\mu_{\text{B}}$. We assume that the displacement of atom M is 1\% of the bond length $a_0$, and the phonon energy is $\hbar\omega=0.02$eV~\cite{Li2019,Bae2022}.} 
		\label{tab:magnetic_moment}
    \begin{tabular}{ccccc} \hline \hline
      & MoS$_2$ & MoSe$_2$ & WS$_2$ & WSe$_2$ \\ \hline    
    $Z^*_{\text{M}}$ &  $-0.09$ & $-0.13$ & $-0.07$ & $-0.12$  \\
    $Z^*_{\text{X}}$ & $0.04$ & $0.04$ & $0.03$ & $0.02$ \\ 
    \hline  
    $L_{\text{ph}}(\hbar)$ &  $1.3$  &  $0.4$ &  $5.3$ &  $2.1$  \\
     $L_{\text{el}}(\hbar)$  &  $6.8\times10^{-6}$  &   $1.4\times10^{-8}$  &  $1.8\times10^{-4}$ &  $2.0\times10^{-6}$  \\ 
    \hline
   $\mu_{\text{ph}}$($\mu_\text{B}$) &  $1.6\times10^{-6}$ & $3.8\times10^{-7}$ & $1.9\times10^{-6}$ & $6.9\times10^{-7}$ \\
   $\mu_{\text{el}}$($\mu_\text{B}$) & $6.8\times10^{-6}$ & $1.4\times10^{-8}$ & $1.8\times10^{-4}$ & $2.0\times10^{-6}$ \\  
    \hline \hline     
      \end{tabular}
\end{table}


\end{document}


\title{Supplemental Material for ``Dynamical Orbital Angular Momentum Induced by Circularly Polarized Phonons"}

\author{Dapeng Yao}
\affiliation{RIKEN Center for Emergent Matter Science (CEMS), 2-1 Hirosawa, Wako, Saitama, 351-0198, Japan}

\author{Dongwook Go}
\affiliation{Department of Physics, Korea University, Seoul 02841, Republic of Korea}
\affiliation{Peter Gr\"{u}nberg Institut and Institute for Advanced Simulation, Forschungszentrum J\"{u}lich and JARA, 52425 J\"{u}lich, Germany}
\affiliation{Institute of Physics, Johannes Gutenberg University Mainz, 55099 Mainz, Germany}

\author{Yuriy Mokrousov}
\affiliation{Peter Gr\"{u}nberg Institut and Institute for Advanced Simulation, Forschungszentrum J\"{u}lich and JARA, 52425 J\"{u}lich, Germany}
\affiliation{Institute of Physics, Johannes Gutenberg University Mainz, 55099 Mainz, Germany}

\author{Shuichi Murakami}
\affiliation{Department of Applied Physics, University of Tokyo, 7-3-1 Hongo, Bunkyo-ku, Tokyo 113-8656, Japan}
\affiliation{International Institute for Sustainability with Knotted Chiral Meta Matter (WPI-SKCM$^\text{2}$),
Hiroshima University, 1-3-1 Kagamiyama, Higashi-Hiroshima, Hiroshima 739-8526, Japan
}
\affiliation{RIKEN Center for Emergent Matter Science (CEMS), 2-1 Hirosawa, Wako, Saitama, 351-0198, Japan}


\maketitle


\renewcommand{\theequation}{S\arabic{equation}}
\setcounter{equation}{0}
\renewcommand{\tablename}{Table S}
\setcounter{table}{0}
\renewcommand{\thefigure}{S\arabic{figure}}
\setcounter{figure}{0}
\renewcommand{\thesection}{\Roman{section}}

\setcounter{section}{0}

\onecolumngrid

\tableofcontents

\clearpage
\section{Derivation of tight-binding Hamiltonian}
\begin{figure}
\begin{center}
\includegraphics[width=12cm]{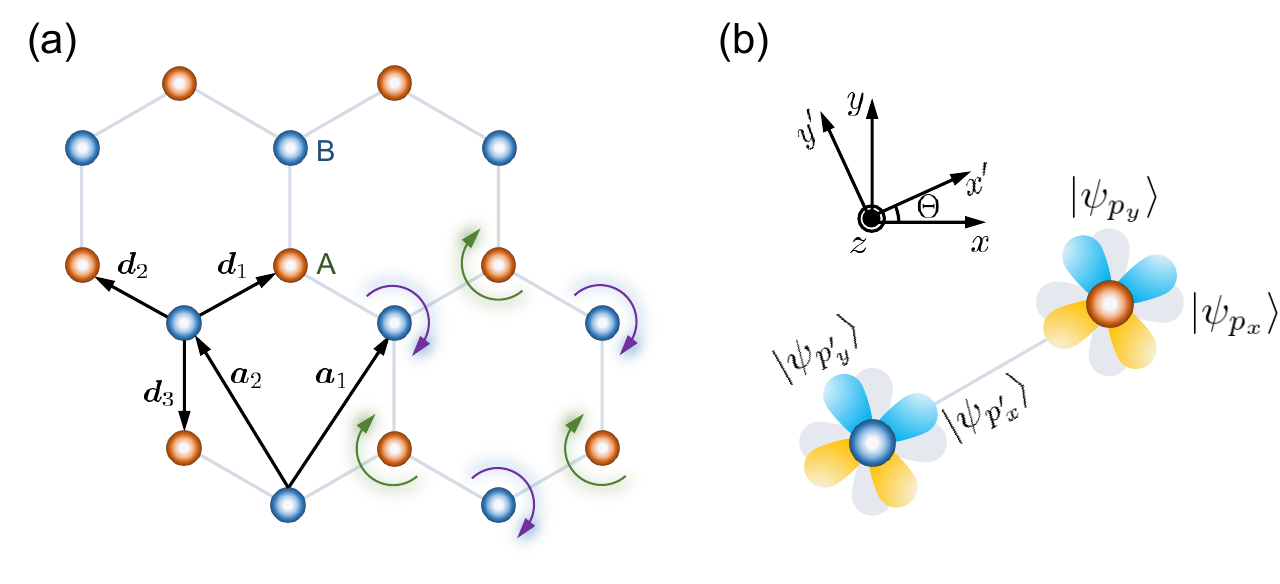}
\caption{Schematic view of the 2D honeycomb lattice model with $p_x$ and $p_y$ orbitals. (a) 2D honeycomb lattice. The bond vectors are $\bm d_{1,2}=(a_0/2)(\pm\sqrt{3},1)$ and $\bm d_{3}=(a_0/2)(0,-1)$ with bond length $a_0=a/\sqrt{3}$.
(b) Rotations of the $p_x$ and $p_y$ orbitals by an angle $\Theta$. Here $\ket{\psi_{p_{x(y)}}}$ and $\ket{\psi_{p'_{x(y)}}}$ represent the orbital wavefunctions corresponding to the $xy$ and $x'y'$ coordinate systems, respectively.}
\label{sm_fig1}
\end{center}
\end{figure}

We provide the detailed derivation of the tight-binding (TB) Hamiltonian modulated by chiral phonons. We consider the 2D honeycomb lattice with $p_x$ and $p_y$ orbitals at A and B sublattice sites as shown in Fig.~\ref{sm_fig2}(a). Let $\hat{c}_{i\alpha}(\hat{c}_{i\alpha}^{\dagger})$ be the annihilation (creation) operator of the electron at the $i$-th site with $p_{\alpha}$ orbitals labeled by $\alpha=x,y$. In the absence of phonons, the electronic TB Hamiltonian is given by
\begin{align}\label{Hamil}
\hat{H}_0=\sum_{\braket{ij}}\sum_{\alpha\beta}t_{ij}^{\alpha\beta}\hat{c}^{\dagger}_{i\alpha}\hat{c}_{j\beta}+\sum_{i}\sum_{\alpha}\Delta\xi_{i}\hat{c}^{\dagger}_{i\alpha}\hat{c}_{i\alpha},
\end{align}
where the first term represents the nearest-neighbor (NN) hopping $t_{ij}^{\alpha\beta}$ given by the Slater-Koster (SK) parameters, and the second term describes a staggered on-site potential $\Delta$ with $\xi_{A(B)}=\pm1$ for the A(B) sublattice.
To calculate the SK parameter, for the hopping between two atoms, we rotate the coordinate axes from $xy$ to $x'y'$ where the $x'$ axis is along the line connecting the two atoms. 
As shown in Fig.~\ref{sm_fig2}(b), we consider a rotation of $p_x$ and $p_y$ orbitals by angle $\Theta$ with respect to $z$ axis as
\begin{align}
\ket{\psi_{p_x}}&=\cos\Theta\ket{\psi_{p_x'}}-\sin\Theta\ket{\psi_{p_y'}}, \\
\ket{\psi_{p_y}}&=\sin\Theta\ket{\psi_{p_x'}}+\cos\Theta\ket{\psi_{p_y'}},
\end{align}
and the overlaps between these orbitals are given by
\begin{align}
\braket{\psi_{p_x}|\hat{H}|\psi_{p_x}}&=V_{pp\sigma}\cos^2\Theta+V_{pp\pi}\sin^2\Theta, \\
\braket{\psi_{p_x}|\hat{H}|\psi_{p_y}}&=V_{pp\sigma}\cos\Theta\sin\Theta-V_{pp\pi}\cos\Theta\sin\Theta, \\
\braket{\psi_{p_y}|\hat{H}|\psi_{p_y}}&=V_{pp\sigma}\sin^2\Theta+V_{pp\pi}\cos^2\Theta,
\end{align}
where $\hat{H}$ represents the the hopping part of the Hamiltonian, and $V_{pp\sigma}=\braket{\psi_{p_x'}|\hat{H}|\psi_{p_x'}}$ and $V_{pp\pi}=\braket{\psi_{p_y'}|\hat{H}|\psi_{p_y'}}$ denote the matrix elements of $\sigma$ and $\pi$ bonds, respectively.
By using the Harrison's method~\cite{Harrison1994}, they can be written as
\begin{align}
V_{pp\sigma}=\frac{\tilde{\eta}_{\sigma}}{d^2},~V_{pp\pi}=\frac{\tilde{\eta}_{\pi}}{d^2},
\end{align}
where $\tilde{\eta}_{\sigma}$ and $\tilde{\eta}_{\pi}$ represent the parameters of $\sigma$ and $\pi$ bonds, respectively, and $d$ denotes the distance between two atoms.
Considering the directions of electron hopping with different $\bm d_{ij}$ and $\Theta_{ij}$, the SK parameters in the TB Hamiltonian Eq.~(\ref{Hamil}) are given by
\begin{align}
t^{xx}_{ij}&=\frac{1}{d^2_{ij}}\left(\tilde{\eta}_{\sigma}\cos^2\Theta_{ij}+\tilde{\eta}_{\pi}\sin^2\Theta_{ij}\right), \\
t^{xy}_{ij}&=\frac{1}{d^2_{ij}}\left(\tilde{\eta}_{\sigma}-\tilde{\eta}_{\pi}\right)\cos\Theta_{ij}\sin\Theta_{ij}, \\
t^{yy}_{ij}&=\frac{1}{d^2_{ij}}\left(\tilde{\eta}_{\sigma}\sin^2\Theta_{ij}+\tilde{\eta}_{\pi}\cos^2\Theta_{ij}\right).
\end{align}
For simplicity, here we consider only the $\sigma$-type hopping $\tilde{\eta}_{\sigma}$. Table~\ref{tab:theta} summarizes these hopping parameters corresponding to the vectors $\bm d_{1,2,3}$ shown in Fig.~\ref{sm_fig1}(a).
Under the basis: $(\ket{\text{A},p_x},\ket{\text{A},p_y},\ket{\text{B},p_x},\ket{\text{B},p_y})$, the Bloch Hamiltonian of Eq.~(\ref{Hamil}) in $\bm k$ space is given by the following matrix:
\begin{align}
\mathcal{H}(\bm k)=
\begin{bmatrix}
\Delta\sigma_0 & h(\bm k) \\
h^{\dagger}(\bm k) & -\Delta\sigma_0
\end{bmatrix},
\end{align}
where 
\begin{align}
h(\bm k)=t_{\sigma}
\begin{bmatrix}
\frac{3}{4}(e^{-i\mathsf{K}_1}+e^{-i\mathsf{K}_2}) & \frac{\sqrt{3}}{4}(e^{-i\mathsf{K}_1}-e^{-i\mathsf{K}_2}) \\
\frac{\sqrt{3}}{4}(e^{-i\mathsf{K}_1}-e^{-i\mathsf{K}_2}) & \frac{1}{4}(e^{-i\mathsf{K}_1}+e^{-i\mathsf{K}_2})+1
\end{bmatrix},
\end{align}
with $\mathsf{K}_{1,2}=\bm k\cdot\bm a_{1,2}$, $\sigma_0$ is a $2\times2$ identity matrix, and $t_{\sigma}=\tilde{\eta}_{\sigma}/a_0^2$.
Here we notice that $\mathcal{H}(\bm k)$ satisfies $C_3$ rotation symmetry with respect to the $z$ axis because of the hexagonal structure, yielding
\begin{align}\label{unitary}
U\mathcal{H}(\bm k)U^{-1}=\mathcal{H}(\hat{R}_3\bm k),
\end{align}
where
\begin{align}
U=
\begin{bmatrix}
1 & \\
& e^{-i\mathsf{K}_2}
\end{bmatrix}\otimes\hat{R}_3,~
\hat{R}_3=
\begin{bmatrix}
-\frac{1}{2} & -\frac{\sqrt{3}}{2} \\
\frac{\sqrt{3}}{2} & -\frac{1}{2}
\end{bmatrix}.
\end{align}
In the absence of $\pi$-type hopping, a TB model on the honeycomb lattice with $p$ orbitals shows two flat bands as shown in the main text, due to the destructive interference~\cite{Wu2007,Wu2008,Liu2014}.

\begin{table}[]
	\centering
	\caption{Table of the angles, vectors, and hopping parameters in the TB Hamiltonian [Eq.~(\ref{Hamil})] for the three directions of NN hoppings along the vectors $\bm d_{\alpha}$ $(\alpha=1,2,3)$.}
		\label{tab:theta}
    \begin{tabular}{cccc} \hline \hline
    $\braket{ij}$ & $\bm d_1=(a_0/2)(\sqrt{3},1)$ & $\bm d_2=(a_0/2)(-\sqrt{3},1)$ & $\bm d_3=a_0(0,-1)$ \\ \hline    
    $\theta_{ij}$ & $\pi/6$ & $5\pi/6$ & $-\pi/2$ \\ 
    $\cos\theta_{ij}$ & $\sqrt{3}/2$ & $-\sqrt{3}/2$ & $0$ \\ 
    $\sin\theta_{ij}$ & $1/2$ & $1/2$ & $-1$ \\ 
    $\bm d_{\alpha}\cdot\bm u$ & $(a_0/2)(\sqrt{3}u_x+u_y)$ & $(a_0/2)(-\sqrt{3}u_x+u_y)$ & $-a_0u_y$
         \\ 
    $t^{xx}_{ij}$ & $3t_{\sigma}/4$ & $3t_{\sigma}/4$ & $0$ \\    
    $t^{yy}_{ij}$ & $t_{\sigma}/4$ & $t_{\sigma}/4$ & $t_{\sigma}$ \\  
    $t^{xy}_{ij}$ & $-\sqrt{3}t_{\sigma}/4$ & $\sqrt{3}t_{\sigma}/4$ & $0$ \\  
    $\delta t^{xx}_{ij}$ & $-t_{\sigma}(\sqrt{3}u_x/2a_0+3u_y/2a_0)$ & $t_{\sigma}(\sqrt{3}u_x/2a_0-3u_y/2a_0)$ & $0$ \\ 
    $\delta t^{yy}_{ij}$ & $-t_{\sigma}(\sqrt{3}u_x/2a_0-u_y/2a_0)$ & $t_{\sigma}(\sqrt{3}u_x/2a_0+u_y/2a_0)$ & $2t_{\sigma}u_y/a_0$ \\ 
    $\delta t^{xy}_{ij}$ & $t_{\sigma}u_x/a_0$ & $t_{\sigma}u_x/a_0$ & $t_{\sigma}u_x/a_0$ \\ \hline  \hline
  \end{tabular}
\end{table}

Next, we introduce chiral phonons into the lattice. For simplicity, we introduce the optical phonons at the Brillouin zone (BZ) center ($\Gamma$ point), in which the mixture of the longitudinal and transverse optical modes can be circularly polarized. In this case, the atoms A and B rotate with a phase difference $\pi$ around their equilibrium positions while the same atom rotates without phase difference as shown in Fig.~\ref{sm_fig1}(a) by arrows. Here we express the phonon displacements for sublattices A and B as $\bm u_A(t)=u_A(\cos{\omega t},\sin{\omega t})$ and $\bm u_B(t)=u_B(\cos{\omega t},\sin{\omega t})$ with phonon frequency $\omega$, and $\bm u(t)=\bm u_B(t)-\bm u_A(t)=u_r(\cos{\omega t},\sin{\omega t})$ represents their relative displacement with $u_r=u_B-u_A$ being the relative amplitude of the atomic rotation.
In this case, the vectors connecting the NN sublattices $\bm d_{ij}$ and the angle $\Theta_{ij}$ for different relative positions are directly modulated by the phonon. First, the distance $\bm d_{ij}$ changes as
\begin{align}
\bm d_{ij}\rightarrow\bm d'_{ij}(t)=\bm d_{ij}+\bm u(t),
\end{align}
leading to
\begin{align}\label{inverse_d2}
\frac{1}{d^2_{ij}}\rightarrow\frac{1}{d^2_{ij}+2\bm d_{ij}\cdot\bm u(t)}\approx\frac{1}{d^2_{ij}}-\frac{2\bm d_{ij}\cdot\bm u(t)}{d^4_{ij}}.
\end{align}
Meanwhile, the angle $\Theta_{ij}$ changes into $\Theta'_{ij}(t)$, which dynamically changes with time $t$, and we have
\begin{align}
\cos\Theta'_{ij}(t)=\frac{d_{ij,x}+u_x(t)}{\sqrt{[d_{ij,x}+u_x(t)]^2+[d_{ij,y}+u_y(t)]^2}},
\sin\Theta'_{ij}(t)=\frac{d_{ij,y}+u_y(t)}{\sqrt{[d_{ij,x}+u_x(t)]^2+[d_{ij,y}+u_y(t)]^2}},
\end{align}
where
\begin{align}
\frac{1}{\sqrt{[d_{ij,x}+u_x(t)]^2+[d_{ij,y}+u_y(t)]^2}}\approx\frac{1}{d_{ij}}\left(1-\frac{\bm d_{ij}\cdot\bm u(t)}{d^2_{ij}}\right).
\end{align}
Thus, the cosine and sine functions of $\Theta_{ij}$ are expanded to the first order of phonon displacement as
\begin{align}
\cos\Theta'_{ij}(t)&\approx\cos\Theta_{ij}+\frac{u_x(t)}{d_{ij}}-\cos\Theta_{ij}\frac{\bm d_{ij}\cdot\bm u(t)}{d^2_{ij}},\label{cos} \\
\sin\Theta'_{ij}(t)&\approx\sin\Theta_{ij}+\frac{u_y(t)}{d_{ij}}-\sin\Theta_{ij}\frac{\bm d_{ij}\cdot\bm u(t)}{d^2_{ij}}. \label{sin}
\end{align}
Combining Eqs.~(\ref{inverse_d2}), (\ref{cos}), and (\ref{sin}), we obtain the modulated hopping amplitudes as follows:
\begin{align}
t^{xx}_{ij}&\rightarrow t^{xx}_{ij}+\delta t^{xx}_{ij}(t), \\
t^{yy}_{ij}&\rightarrow t^{yy}_{ij}+\delta t^{yy}_{ij}(t), \\
t^{xy}_{ij}&\rightarrow t^{xy}_{ij}+\delta t^{xy}_{ij}(t), 
\end{align}
where
\begin{align}
\delta t^{xx}_{ij}(t)=&\frac{2t_{\sigma}}{a_0}\cos\Theta_{ij}u_x(t)-\frac{4t_{\sigma}}{a_0^2}\cos^2\Theta_{ij}\bm d_{ij}\cdot\bm u(t), \\
\delta t^{yy}_{ij}(t)=&\frac{2t_{\sigma}}{a_0}\sin\Theta_{ij}u_y(t)-\frac{4t_{\sigma}}{a_0^2}\sin^2\Theta_{ij}\bm d_{ij}\cdot\bm u(t), \\
\delta t^{xy}_{ij}(t)=&\frac{t_{\sigma}}{a_0}\left[u_x(t)\sin\Theta_{ij}+u_y(t)\cos\Theta_{ij}\right]-\frac{4t_{\sigma}}{a_0^2}\cos\Theta_{ij}\sin\Theta_{ij}\bm d_{ij}\cdot\bm u(t),
\end{align}
where we used $d_{ij}=a_0$. Here the modulated hopping parameter $\delta t_{ij}^{\alpha\beta}\sim (u_r/a_0)t_{ij}^{\alpha\beta}$ is of the first order in the relative amplitude.
These parameters are summarized into Table~\ref{tab:theta} for the vectors $\bm d_{1,2,3}$ as shown in Fig.~\ref{sm_fig1}(a).
As a result, the modulated TB Hamiltonian is given by
\begin{align}
\delta\hat{H}(t)=\sum_{\braket{ij}}\sum_{\alpha\beta}\delta t_{ij}^{\alpha\beta}(t)\hat{c}_{i\alpha}^{\dagger}\hat{c}_{j\beta},
\end{align}
which stands for electron-phonon coupling, and the total Hamiltonian is $\hat{H}(t)=\hat{H}_0+\delta\hat{H}(t)$.
The Bloch Hamiltonian of $\delta\hat{H}(t)$ in $\bm k$ space is given by
\begin{align}
\delta\mathcal{H}(\bm k,t)=
\begin{bmatrix}
 & \delta h(\bm k,t) \\
\delta h^{\dagger}(\bm k,t) & 
\end{bmatrix},
\end{align}
where
\begin{align}
&\delta h(\bm k,t)=-\frac{u_rt_{\sigma}}{a_0}\times\nonumber\\
&\begin{bmatrix}
\frac{\sqrt{3}u_x(t)}{2}\left(e^{-i\mathsf{K}_1}-e^{-i\mathsf{K}_2}\right)+\frac{3u_y(t)}{2}\left(e^{-i\mathsf{K}_1}+e^{-i\mathsf{K}_2}\right) & u_x(t)\left(1+e^{-i\mathsf{K}_1}+e^{-i\mathsf{K}_2}\right) \\
u_x(t)\left(1+e^{-i\mathsf{K}_1}+e^{-i\mathsf{K}_2}\right) & \frac{\sqrt{3}u_x(t)}{2}\left(e^{-i\mathsf{K}_1}-e^{-i\mathsf{K}_2}\right)-\frac{u_y(t)}{2}\left(e^{-i\mathsf{K}_1}+e^{-i\mathsf{K}_2}\right)-2u_y(t)
\end{bmatrix}.
\end{align}
We can check that the total Bloch Hamiltonian $\mathcal{H}(\bm k,t)=\mathcal{H}(\bm k,\bm u)$, with $\bm u=\bm u(t)$, respects $C_3$ rotation symmetry, which is represented by the following unitary transformation:
\begin{align}
U\mathcal{H}(\bm k,\bm u)U^{-1}=\mathcal{H}(\hat{R}_3\bm k,\hat{R}_3\bm u),
\end{align}
where $U$ is given by Eq.~(\ref{unitary}).

\section{Calculation of phonon dispersion}
We show the details of phonon dispersion calculation in the main text.
The general phonon Lagrangian with time-reversal symmetry in the harmonic approximation is given by
\begin{align}
\mathcal{L}=\frac{1}{2}\sum_{l\kappa}m_{\kappa}\dot{\bm u}^2_{l\kappa}-\frac{1}{2}\sum_{ll'\kappa\kappa'}\sum_{\alpha,\beta=x,y,z}\Phi_{\alpha\beta}(l\kappa,l'\kappa')u_{l\kappa,\alpha}u_{l'\kappa',\beta},
\end{align}
where $\bm u_{l\kappa}$ is the displacement vector of the $\kappa$th atom in the $l$th unit cell, $m_{\kappa}$ is the mass of the $\kappa$th atom in the unit cell, and $\Phi$ is the force constant matrix~\cite{Hamada2020}. The equation of motion is obtained from the Euler-Lagrange equation as
\begin{align}\label{EOM_ph}
m_{\kappa}\ddot{u}_{l\kappa,\alpha}=-\sum_{l'\kappa'\beta}\Phi_{\alpha\beta}(l\kappa,l'\kappa')u_{l'\kappa'\beta}.
\end{align}
Then we introduce the Fourier transformation of the displacement vector as
\begin{align}
\sqrt{m_{\kappa}}\bm u_{l\kappa}=\frac{1}{\sqrt{N}}\sum_{\textbf k}U_{\textbf k\kappa}e^{i\textbf R_{l}\cdot\textbf{k}-i\omega t},
\end{align}
and Eq.~(\ref{EOM_ph}) is rewritten by
\begin{align}\label{eigenvalue_ph}
\omega^2U_{\textbf k\kappa,\alpha}=\sum_{\kappa'\beta}D_{\alpha\beta}(\textbf k,\kappa\kappa')U_{\textbf k\kappa',\beta}
\end{align}
with
\begin{align}
D_{\alpha\beta}(\textbf k,\kappa\kappa')=\sum_{l'}\frac{\Phi_{\alpha\beta}(l\kappa,l'\kappa')}{\sqrt{m_{\kappa}m_{\kappa'}}}e^{i\textbf k\cdot(\textbf R_{l'}-\textbf R_{l})}
\end{align}
being the Hermitian dynamical matrix: $D_{\alpha\beta}(\textbf k,\kappa\kappa')=D_{\beta\alpha}^*(\textbf k,\kappa'\kappa)$. Thus, the dynamical matrix is a $3N\times3N$ Hermitian matrix, and its eigenvalues $\omega^2_{\textbf k\sigma}$ are real. Solving Eq.~(\ref{eigenvalue_ph}) reduces to the following eigenvalue problem:
\begin{align}\label{eigenvalue_ph_final}
D(\textbf k)\bm\epsilon_{\textbf k\sigma}=\omega_{\textbf k\sigma}^2\bm\epsilon_{\textbf k\sigma},
\end{align}
where $\bm\epsilon_{\textbf k\sigma}$ is the displacement polarization vector including all the phonon modes.

\begin{figure}
\begin{center}
\includegraphics[width=12cm]{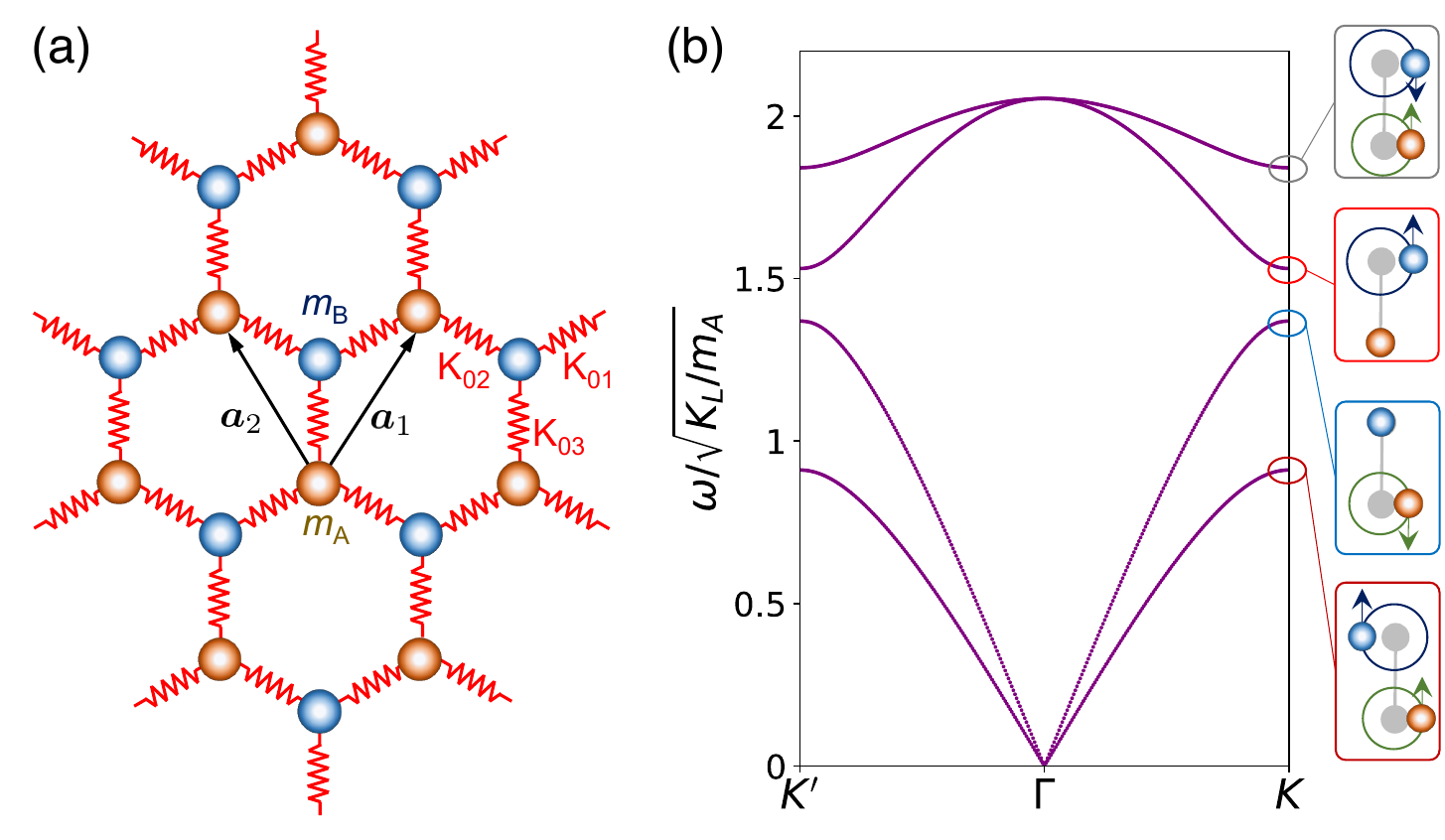}
\caption{(a) Schematic picture of the spring-mass model on the 2D honeycomb lattice. Each unit cell has two atoms with masses $m_A$ and $m_B$. The couplings between the nearest-neighboring atoms are represented by $K_{01}$, $K_{02}$, and $K_{03}$, which are $2\times2$ matrices. The primitive translation vectors are given by $\bm a_1=a(1/2,\sqrt{3}/2)$ and $\bm a_2=a(-1/2,\sqrt{3}/2)$. (b) Phonon dispersion and the circularly polarized modes at the $K$ point with $m_B=0.8m_A$ and $K_T=K_L/4$. We also show the displacement of the atoms.}
\label{sm_fig2}
\end{center}
\end{figure}

We apply Eq.~(\ref{eigenvalue_ph_final}) to a spring-mass model forming the two-dimensional (2D) honeycomb lattice as shown in Fig.~\ref{sm_fig1}(a). Each unit cell has two atoms with masses $m_A$ and $m_B$. The primitive translation vectors are given by $\bm a_1=a(1/2,\sqrt{3}/2)$ and $\bm a_2=a(-1/2,\sqrt{3}/2)$ with lattice constant $a$.
Here we discuss the 2D motion within the $xy$ plane and consider springs between the NN atoms. We let the spring constant matrix along the $x$ axis be
\begin{align}
K_x=\begin{bmatrix}
K_L & \\
& K_T
\end{bmatrix},
\end{align}
which is constructed from the longitudinal and transverse spring constants $K_T$ and $K_L$, respectively, corresponding to the deformations perpendicular and parallel to the spring~\cite{Zhang2015}.
By defining a rotation operator within the $xy$ plane as
\begin{align}
R(\theta)=\begin{bmatrix}
\cos{\theta} & -\sin{\theta} \\
\sin{\theta} & \cos{\theta}
\end{bmatrix},
\end{align}
we can obtain the three spring constants within the unit cell:
\begin{align}
K_{01}=R(\frac{\pi}{6})K_xR(-\frac{\pi}{6}),\\
K_{02}=R(-\frac{\pi}{6})K_xR(\frac{\pi}{6}),\\
K_{03}=R(-\frac{\pi}{2})K_xR(\frac{\pi}{2}).
\end{align}
Under the basis: $\bm\epsilon_{\textbf k}=(\sqrt{m_A}\epsilon_{\textbf kAx},\sqrt{m_A}\epsilon_{\textbf kAy},\sqrt{m_B}\epsilon_{\textbf kBx},\sqrt{m_B}\epsilon_{\textbf kBy})^{\mathsf{T}}$ with $\epsilon_{\textbf k(A,B)(x,y)}$ being the $x(y)$-component displacement of the atom A(B), the $4\times4$ dynamical matrix is written as
\begin{align}
D(\textbf k)=K_0+K_1e^{i\textbf k\cdot\bm a_1}+K_2e^{-i\textbf k\cdot\bm a_2}+K_3e^{i\textbf k\cdot\bm a_2}+K_4e^{i\textbf k\cdot\bm a_1},
\end{align}
where
\begin{align}
K_0&=\begin{bmatrix}
\frac{K_{01}+K_{02}+K_{03}}{m_A} & -\frac{K_{03}}{\sqrt{m_Am_B}} \\
-\frac{K_{03}}{\sqrt{m_Am_B}} & \frac{K_{01}+K_{02}+K_{03}}{m_B}
\end{bmatrix},\\
K_1&=\begin{bmatrix}
 0 & 0 \\
-\frac{K_{01}}{\sqrt{m_Am_B}} & 0
\end{bmatrix},
K_2=\begin{bmatrix}
 0 & 0 \\
-\frac{K_{02}}{\sqrt{m_Am_B}} & 0
\end{bmatrix},\\
K_3&=\begin{bmatrix}
0 & -\frac{K_{02}}{\sqrt{m_Am_B}} \\
0 & 0
\end{bmatrix},
K_4=\begin{bmatrix}
0 & -\frac{K_{01}}{\sqrt{m_Am_B}} \\
0 & 0
\end{bmatrix}.
\end{align}
with $K_1$, $K_2$, $K_3$, and $K_4$ being the couplings between the adjacent unit cells.
The phonon dispersion and four circularly polarized modes at the $K$ point are shown in Fig.~\ref{sm_fig2}(b) with $m_B=0.8m_A$.

\section{Orbital angular momentum (OAM)}\label{III}
\subsection{Dynamical OAM}
We illustrate the calculation details on the OAM induced by chiral phonons.
In an adiabatic approximation, we consider a band insulator and assume that the phonon frequency $\omega$ is much lower than the electronic band gap.
In this case, an expectation value of a physical quantity $\hat{X}$ during a time period $T$ driven by phonons is given by
\begin{align}\label{Lz_geom}
\bar{X}=&\frac{1}{V}\int_0^Tdt\sum_n\left[\bra{\psi_n(t)}\hat{X}\ket{\psi_n(t)}+\sum_{m(\neq n)}\frac{\hbar\bra{\psi_n(t)}\hat{X}\ket{\psi_m(t)}\bra{\psi_m(t)}i\partial_t\ket{\psi_n(t)}}{E_n(t)-E_m(t)}+\text{c.c}\right]\nonumber\\
=&\frac{1}{2\pi}\int_0^{2\pi}d\tau\sum_n\left[\bra{\psi_n(\tau)}\hat{X}\ket{\psi_n(\tau)}+\omega\sum_{m(\neq n)}\frac{\hbar\bra{\psi_n(\tau)}\hat{X}\ket{\psi_m(\tau)}\bra{\psi_m(\tau)}i\partial_{\tau}\ket{\psi_n(\tau)}}{E_n(\tau)-E_m(\tau)}+\text{c.c}\right],
\end{align}
where we replace the time $t$ by a dimensionless quantity $\tau=\omega t$ for convenience~\cite{HamadaGeo2020}.
Here the first term is the instantaneous term, which represents the snapshot of the eigenstates $\ket{\psi_n(t)}$ averaged over one period. It vanishes in nonmagnetic electronic systems due to the time-reversal symmetry. The second term is the geometric term, which is essential for the emergence of OAM (let $\hat{X}=\hat{\bm L}$). 
The geometric term of the OAM for the $n$th band is given by
\begin{align}
\bar{L}_{z,n}=\frac{1}{2\pi}\int_0^{2\pi}d\tau\sum_{m\neq n}\left[\frac{\hbar\omega\hat{L}_{z,mn}(\tau)A_{nm}(\tau)}{E_n(\tau)-E_m(\tau)}+\text{c.c}\right],
\end{align}
where $\hat{L}_{z,mn}(\tau)=\bra{\psi_n(t)}\hat{L}_z\ket{\psi_m(t)}$ is the matrix element of $\hat{L}_z$ and $A_{nm}(\tau)=\bra{\psi_m(t)}i\partial_{\tau}\ket{\psi_n(t)}$ is the Berry phase.
When either counterclockwise (CCW) or clockwise (CW) phonon mode is present, the geometric term is nonzero because of the dynamically acquired Berry phase of electrons.
The geometric OAM is proportional to the phonon frequency $\omega$.
We notice that the sign of the time-averaged OAM is opposite between CCW and CW phonons, because the chiral phonon with CCW and CW modes are mutually related via the time-reversal symmetry (TRS): $\tau\rightarrow -\tau$. Namely, the matrix element of $\hat{L}_z$ and Berry connection are connected by $L^{\text{CCW}}_{z,nm}(\tau)=L^{\text{CW}}_{z,nm}(-\tau)$ and $A^{\text{CCW}}_{mn}(\tau)=-A^{\text{CW}}_{mn}(-\tau)$, yielding $L^{\text{CCW}}_{z}(\tau)=-L^{\text{CW}}_{z}(-\tau)$.

\subsection{Time-averaged OAM}
In addition to the time-dependent expression of OAM, here we further derive its time average in the similar way as in the spin angular momentum~\cite{Yao2025}. We first introduce the electronic orbital magnetic moment $\hat{\bm\mu}_L=-\frac{e}{2m}\hat{\bm L}=-\frac{\mu_B}{\hbar}\hat{\bm L}$, and an orbital Zeeman field $\bm B$ as a conjugate filed to $\hat{\bm\mu}_L$. The Hamiltonian is given by
\begin{align}
\hat{H}_{B}=-\hat{\bm\mu}_L\cdot\bm B=\frac{\mu_B}{\hbar}\hat{\bm L}\cdot\bm B,
\end{align}
where $\mu_B$ is the Bohr magneton, and $\hat{\bm L}$ is the electronic OAM operator. In the absence of the Zeeman field, the $i$ $(i=x,y,z)$ component of the OAM is given by
\begin{align}\label{LiBH}
\hat{L}_{i}=\frac{\hbar}{\mu_B}\partial_{B_{i}}\hat{H}\Big|_{\bm B=0},
\end{align}
with the total Hamiltonian $\hat{H}$. From Eq.~(\ref{Lz_geom}), the dynamical OAM for the $n$th electronic band from the geometric effect at time $t$ is
\begin{align}\label{Lint}
L_{i,n}(t)=\int\frac{d^dk}{(2\pi)^d}\sum_{m(\neq n)}\left[\frac{\hbar\hat{L}_{i,nm}(t)A_{mn}(t)}{E_n(t)-E_m(t)}+\text{c.c}\right],
\end{align}
with the matrix element of OAM $\hat{L}_{i,nm}(t)=\bra{\psi_n(t)}\hat{L}_{i}\ket{\psi_m(t)}$ and the Berry connection $A_{mn}(t)=i\bra{\psi_m(t)}\partial_t\ket{\psi_n(t)}$. Here the matrix element of OAM is further expressed as
\begin{align}\label{Linm}
\hat{L}_{i,nm}(t)=\frac{\hbar}{\mu_B}\bra{\psi_n(t)}\partial_{B_i}\hat{H}\big|_{\bm B=0}\ket{\psi_m(t)}=-\frac{\hbar}{\mu_B}[E_n(t)-E_m(t)]\braket{\psi_{n}(t)|\partial_{B_i}\psi_m(t)}\big|_{\bm B=0}.
\end{align}

We assume that the crystal consists of only one species of atoms, and an atomic motion is in-plane with the displacement $\bm u=(u_x,u_y)$ of one representative atom periodically changing with time~\cite{Ren2021,Yao2025}, and then we can replace the Berry connection defined by the time derivative by the one of displacement derivative as
\begin{align}\label{Amn}
A_{mn}(\bm u)=i\braket{\psi_m(\bm u)|\nabla_{\bm u}\psi_n(\bm u)}\cdot\dot{\bm u}.
\end{align}
By substituting Eqs.~(\ref{Linm}) and (\ref{Amn}) into (\ref{Lint}), we can express the OAM as
\begin{align}\label{LinB}
L_{i,n}&=\frac{\hbar^2}{\mu_B}\sum_{\delta=x,y}\dot{u}_{\delta}\int\frac{d^dk}{(2\pi)^d}\left[i\braket{\partial_{B_i}\psi_n|\partial_{u_{\delta}}\psi_n}+\text{c.c}\right]\big|_{\bm B=0}\nonumber\\
&=\frac{\hbar^2}{\mu_B}\sum_{\delta=x,y}\dot{u}_{\delta}\int\frac{d^dk}{(2\pi)^d}\Omega^{(n)}_{B_{i}u_{\delta}}\Big|_{\bm B=0},
\end{align}
where $\Omega^{(n)}_{B_{i}u_{\delta}}=\partial_{B_i}A^{(n)}_{u_{\delta}}-\partial_{u_{\delta}}A^{(n)}_{B_i}$ is the Berry curvature with $A^{(n)}_{\rho}=i\braket{\psi_n|\partial_{\rho}\psi_n}$ $(\rho=u_{\delta},B_{i})$ being the Berry connection in a general form. Since the displacement $u_{\delta}$ is smaller compared with the lattice constant, Eq.~(\ref{LinB}) can be expended near $\bm u=0$ to the first order of $\bm u$ as
\begin{align}\label{Liapp}
L_{i,n}\approx\frac{\hbar^2}{\mu_B}\sum_{\delta}\dot{u}_{\delta}\int\frac{d^dk}{(2\pi)^d}\Omega^{(n)}_{B_{i}u_{\delta}}\Big|_{\bm u=0,\bm B=0}+\frac{\hbar^2}{\mu_B}\sum_{\delta\gamma}\dot{u}_{\delta}u_{\gamma}\int\frac{d^dk}{(2\pi)^d}\partial_{u_{\gamma}}\Omega^{(n)}_{B_iu_{\delta}}\Big|_{\bm u=0,\bm B=0},
\end{align}
where the Berry curvatures are evaluated at $\bm u=0$. In particular, the following term can be separated into symmetric and antisymmetric parts as
\begin{align}
\sum_{\delta\gamma}\dot{u}_{\delta}u_{\gamma}\partial_{u_{\gamma}}\Omega^{(n)}_{B_iu_{\delta}}=&~u_x\dot{u}_x\partial_{u_x}\Omega^{(n)}_{B_iu_x}+u_y\dot{u}_y\partial_{u_y}\Omega^{(n)}_{B_iu_y}\nonumber\\
&+\frac{1}{2}(u_x\dot u_y+u_y\dot u_x)\left(\partial_{u_x}\Omega^{(n)}_{B_iu_y}+\partial_{u_y}\Omega^{(n)}_{B_iu_x}\right)\nonumber\\
&+\frac{1}{2}(u_x\dot u_y-u_y\dot u_x)\left(\partial_{u_x}\Omega^{(n)}_{B_iu_y}-\partial_{u_y}\Omega^{(n)}_{B_iu_x}\right).
\end{align}
After taking the time average over the time period $T=2\pi/\omega$, the first term in Eq.~(\ref{Liapp}) and the symmetric part in the second term vanish because of the periodicity of $u_{\delta}$, and only the antisymmetric part contributes to the time-averaged OAM as
\begin{align}\label{Linfinal}
\bar{L}_{i,n}=\frac{\hbar^2}{2\mu_B}J_z^{\text{ph}}\int\frac{d^dk}{(2\pi)^d}\partial_{B_i}\Omega^{(n)}_{u_xu_y}\Big|_{\bm u=0,\bm B=0},
\end{align}
where $J_z^{\text{ph}}=\int_0^Tdt(\bm u\times\dot{\bm u})_z$ represents the phonon angular momentum divided by atomic mass, and $\Omega^{(n)}_{u_xu_y}=\partial_{u_x}A^{(n)}_{u_y}-\partial_{u_y}A^{(n)}_{u_x}$ denotes the Berry curvature defined in phonon displacement space coming from the identity $\partial_{u_x}\Omega^{(n)}_{B_iu_y}-\partial_{u_y}\Omega^{(n)}_{B_iu_x}=\partial_{B_i}\Omega^{(n)}_{u_xu_y}$.

Since the $B_i$ derivative appearing in Eq.~(\ref{Linfinal}) is analytically intractable in the absence of the orbital Zeeman field, we further derive the formula without calculating this derivative. The Berry curvature for the $n$th occupied band is given by
\begin{align}
\Omega^{(n)}_{u_xu_y}=\sum_{m(\neq n)}\left[\frac{iX^{nm}_{u_x}X^{mn}_{u_y}}{(E_n-E_m)^2}+\text{c.c}\right],
\end{align}
with $X_{u_{\delta}}^{nm}\equiv\bra{\psi_n}\partial_{u_{\delta}}\hat{H}\ket{\psi_m}$. Then its $B_i$ derivative reads
\begin{align}\label{partB}
\partial_{B_{i}}\Omega_{u_xu_y}^{(n)}=\sum_{m}^{\text{unocc}}\Bigg[i\frac{(\partial_{B_{i}}X_{u_x}^{nm})X_{u_y}^{mn}+X_{u_x}^{nm}(\partial_{B_{i}}X_{u_y}^{mn})}{(E_n-E_m)^2}-2i\frac{X_{u_x}^{nm}X_{u_y}^{mn}(\partial_{B_{i}}E_n-\partial_{B_{i}}E_m)}{(E_n-E_m)^3}+\text{c.c}\Bigg],
\end{align}
where we replace the Zeeman field derivative of the Hamiltonian by $\partial_{B_i}\hat{H}=\frac{\hbar}{\mu_B}\hat{L}_i$ [see Eq.~(\ref{LiBH})] as follows:
\begin{align}
\partial_{B_{i}}X_{u_{\delta}}^{nm}=&\bra{\partial_{B_{i}}\psi_n}\partial_{u_{\delta}}\hat{H}\ket{\psi_m}+\bra{\psi_n}\partial_{u_{\delta}}\hat{H}\ket{\partial_{B_{i}}\psi_m}+\bra{\psi_n}\partial_{B_{i}}\partial_{u_{\delta}}\hat{H}\ket{\psi_m} \nonumber\\
=&\sum_{l(\neq n)}\frac{\bra{\psi_n}\partial_{B_{i}}\hat{H}\ket{\psi_l}\bra{\psi_l}\partial_{u_{\delta}}\hat{H}\ket{\psi_m}}{E_n-E_l}+\sum_{l(\neq m)}\frac{\bra{\psi_n}\partial_{u_{\delta}}\hat{H}\ket{\psi_l}\bra{\psi_l}\partial_{B_{i}}\hat{H}\ket{\psi_m}}{E_m-E_l} \nonumber\\
=&~\frac{\mu_B}{\hbar}\sum_{l(\neq n)}\frac{\hat{L}^{nl}_{i}X^{lm}_{u_{\delta}}}{E_n-E_l}+\frac{\mu_B}{\hbar}\sum_{l(\neq n)}\frac{X^{nl}_{u_{\delta}}\hat{L}^{lm}_{i}}{E_m-E_l},\\
\partial_{B_{i}}E_n=&\bra{\psi_n}\partial_{B_{i}}\hat{H}\ket{\psi_n}=\frac{\mu_B}{\hbar}\hat{L}^{nn}_i,
\end{align}
with $\delta=x,y$. Therefore, we obtain the expression of the derivative of the Berry curvature as
\begin{align}
\sum_n^{\text{occ}}\partial_{B_{\alpha}}\Omega^{(n)}_{u_xu_y}=&~i\frac{\mu_B}{\hbar}\sum_n^{\text{occ}}\sum_m^{\text{unocc}}\sum_{l}^{\text{unocc}}\frac{\hat{L}_{\alpha}^{nl}X_{u_x}^{lm}X_{u_y}^{mn}+X_{u_x}^{nm}X_{u_y}^{ml}\hat{L}_{\alpha}^{ln}}{(E_n-E_m)^2(E_n-E_l)}\nonumber\\
&+i\frac{\mu_B}{\hbar}\sum_n^{\text{occ}}\sum_m^{\text{unocc}}\sum_{l}^{\text{occ}}\frac{X_{u_x}^{nl}\hat{L}_{\alpha}^{lm}X_{u_y}^{mn}+X_{u_x}^{nm}\hat{L}_{\alpha}^{ml}X_{u_y}^{ln}}{(E_n-E_m)^2(E_m-E_l)}\nonumber\\
&-2i\frac{\mu_B}{\hbar}\sum_n^{\text{occ}}\sum_{m}^{\text{unocc}}\frac{X_{u_x}^{nm}X_{u_y}^{mn}}{(E_n-E_m)^3}\left(\hat{L}_{\alpha}^{nn}-\hat{L}_{\alpha}^{mm}\right)+\text{c.c}.
\end{align}
Here we clearly see that the time-averaged OAM shows an inverse cubic dependence on the energy gap, which is in agreement with our model calculation results in the main text.

\section{Effect of phonons at valley points}
To study the chiral phonons at valley ($K$ or $K'$) points, where the circularly polarized modes appear with a phase difference $e^{\pm i2\pi/3}$ between the same species of atoms, here we construct an effective electronic Hamiltonian with $p$ orbitals with chiral phonons at valley points.

\subsection{Effective Hamiltonian}
\begin{figure}
\begin{center}
\includegraphics[width=15cm]{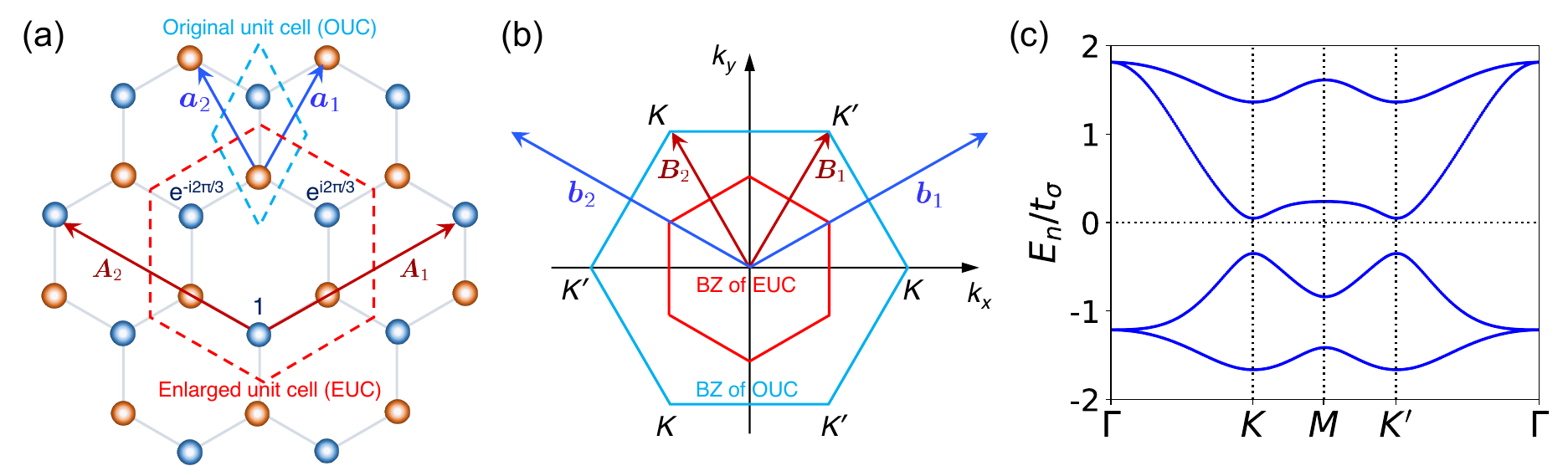}
\caption{Valley ($K$ or $K'$) chiral phonons can be treated as $\Gamma$-chiral phonons by enlarging the unit cell. (a) Chiral phonons at the $K$ point on 2D honeycomb lattice with a phase difference $e^{\pm i2\pi/3}$ between the same species of atoms in different unit cells. The EUC corresponding to this phase factor is enclosed by the red dashed. (b) The BZ of the EUC is reduced from that of OUC. (c) Band structure for the OUC including the NNN hopping term with the parameter $t'_{\sigma}=0.1t_{\sigma}$.}
\label{sm_fig3}
\end{center}
\end{figure}

To calculate the electronic states with chiral phonons at valley points, we enlarge the unit cell by three times, changing the original unit cell (OUC) into the enlarged unit cell (EUC) as shown in Fig.~\ref{sm_fig3}(a). The primitive translation vectors of the OUC: $\bm a_1=a/2(1,\sqrt{3})$ and $\bm a_2=a/2(-1,\sqrt{3})$ are replaced by those of the EUC: $\bm A_1=a/2(3,\sqrt{3})$ and $\bm A_2=a/2(-3,\sqrt{3})$, by which the first BZ of the OUC is reduced into that of EUC as shown in Fig.~\ref{sm_fig3}(b), and the $K$ and $K'$ points in the BZ of the OUC go to the $\Gamma$ point in the EUC.
Here, for simplicity, we introduce an effective Hamiltonian as a minimal model describing the electronic orbital effect induced by valley phonons.

In order to check the orbitals hybridization of electrons at valley points, we first calculate the electronic TB Hamiltonian in Eq.~(\ref{Hamil}). Since the phonons at valley points directly modulate the next-nearest-neighbor (NNN) electron hoppings, we also consider the following NNN hopping term:
\begin{align}
\hat{H}_{\text{NNN}}=\sum_{\braket{\braket{ij}}}\sum_{\alpha\beta}(t'_{ij})^{\alpha\beta}\hat{c}^{\dagger}_{i\alpha}\hat{c}_{j\beta},
\end{align}
where the orbital overlaps between in the NNN term are
\begin{align}\label{H_NNN}
(t'_{ij})^{xx}&=t'_{\sigma}\cos^2\Theta_{ij},\\
(t'_{ij})^{yy}&=t'_{\sigma}\sin^2\Theta_{ij},\\
(t'_{ij})^{xy}&=t'_{\sigma}\cos\Theta_{ij}\sin\Theta_{ij},
\end{align}
with parameter $t'_{\sigma}$. Under the basis: $(\ket{\text{A},p_x},\ket{\text{A},p_y},\ket{\text{B},p_x},\ket{\text{B},p_y})$, the Bloch Hamiltonian of Eq.~(\ref{H_NNN}) is given by
\begin{align}\label{Hk_NNN}
\mathcal{H}_{\text{NNN}}(\bm k)=t'_{\sigma}\sigma_0\otimes\begin{bmatrix}
2\cos{\left(\mathsf{K}_1-\mathsf{K}_2\right)}+\frac{1}{2}\left(\cos{\mathsf{K}_1}+\cos{\mathsf{K}_2}\right) & \frac{\sqrt{3}}{2}\left(\cos{\mathsf{K}_1}-\cos{\mathsf{K}_2}\right) \\
\frac{\sqrt{3}}{2}\left(\cos{\mathsf{K}_1}-\cos{\mathsf{K}_2}\right) & \frac{3}{2}\left(\cos{\mathsf{K}_1}+\cos{\mathsf{K}_2}\right)
\end{bmatrix},
\end{align}
and the band structure is shown in Fig.~\ref{sm_fig3}(c).
Here we focus on the eigenstates at the $K$ and $K'$ points, where the $p_x$ and $p_y$ orbitals hybridize with each other, resulting in four states: $\ket{K_A,p_-}$ and $\ket{K'_B,p_-}$ below $E=0$, $\ket{K_B,p_+}$ and $\ket{K'_A,p_+}$ above $E=0$ after diagonalizing the total TB Hamiltonian including Eq.~(\ref{Hk_NNN}).
Therefore, in the absence of phonons, we consider the electronic TB Hamiltonian with $p_{\pm}=p_x\pm ip_y$ orbitals given by
\begin{align}\label{H0ab}
\hat{H}^{\text{EUC}}_0&=\sum_{\braket{ij}}\sum_{\alpha\beta=\pm}t_{ij}^{\alpha\beta}\hat{c}^{\dagger}_{i\alpha}\hat{c}_{j\beta}+\sum_{\braket{\braket{ij}}}\sum_{\alpha\beta=\pm}(t'_{ij})^{\alpha\beta}\hat{c}^{\dagger}_{i\alpha}\hat{c}_{j\beta}+\sum_{i\alpha=\pm}\Delta\xi_i\hat{c}^{\dagger}_{i\alpha}\hat{c}_{i\alpha}.
\end{align}

We then consider that the chiral phonons at the valley points modulate the Hamiltonian in Eq.~(\ref{H0ab}), leading to a time-dependent term: 
\begin{align}\label{dH0ab}
\delta\hat{H}^{\text{EUC}}(t)&=-\sum_{\braket{ij}}\sum_{\alpha\beta=\pm}t_{ij}^{\alpha\beta}\frac{2\bm d_{ij}\cdot\bm u_{ij}(t)}{a^2_0}\hat{c}^{\dagger}_{i\alpha}\hat{c}_{j\beta}-\sum_{\braket{\braket{ij}}}\sum_{\alpha\beta=\pm}(t'_{ij})^{\alpha\beta}\frac{2\bm a_{ij}\cdot\bm u_{ij}(t)}{a^2}\hat{c}^{\dagger}_{i\alpha}\hat{c}_{j\beta},
\end{align}
where the hopping parameters are given by
\begin{align}
t_{ij}=t_{\sigma}\begin{bmatrix}
1 & e^{-i2\Theta_{ij}} \\
e^{i2\Theta_{ij}} & 1
\end{bmatrix},
t'_{ij}=t'_{\sigma}\begin{bmatrix}
1 & e^{-i2\Theta_{ij}} \\
e^{i2\Theta_{ij}} & 1
\end{bmatrix}.
\end{align}
The effective Hamiltonian under the basis: $(\ket{K_A,p_-}, \ket{K_B,p_-}, \ket{K'_A,p_+},\ket{K'_B,p_-})$ can be obtained by calculating the TB Hamiltonian including both Eqs.~(\ref{H0ab}) and~(\ref{dH0ab}) under the EUC. We show the detailed forms of the effective Hamiltonian in the main text. The intravalley-scattering terms between $K_A$-$K_B$ and $K'_A$-$K'_B$ appear due to the NN hopping between atoms A and B in Eq.~(\ref{H0ab}), while the intervalley-scattering terms between $K_A$-$K'_B$, $K_B$-$K_A'$, $K_A$-$K'_A$, and $K_B$-$K'_B$ appear due to the electron-phonon couling in Eq.~(\ref{dH0ab}), which take different forms classified by different phonon pseudo-angular momentum (PAM) as we discussed in the main text.

\subsection{OAM calculated from the effective Hamiltonian}
In the main text, we show the OAM induced by chiral phonons with different PAM. In our effective Hamiltonian with $p$ orbitals, we notice that the OAM induced by the chiral phonon at $K$ point with $l_{\text{ph}}=1$ is zero as shown in the main text. Here we analyze the reason of the vanishing OAM. In the case of $l_{\text{ph}}=1$, the effective Hamiltonian takes the form as $\mathcal{H}_{\text{eff}}^{(l_{\text{ph}}=1)}=\bm R\cdot\bm\Gamma$, where $\bm R=(\Delta,-\hbar v_Fq_x,\hbar v_Fq_y,-\lambda(u_y-v_y),-\lambda(u_x+v_x))$, and $\bm\Gamma=(s_0\otimes\sigma_z,s_z\otimes\sigma_x,s_0\otimes\sigma_y,s_x\otimes \sigma_x,s_y\otimes\sigma_x)$ with the Fermi velocity $v_F=3a_0t_{\sigma}/2\hbar$, the parameter $\lambda=-3t_{\sigma}/a_0$, $(u_x,u_y)$ and $(v_x,v_y)$ are the displacements of atom A and B.
In this case, the effective Hamiltonian preserves the following two symmetries:
\begin{align}
&\Gamma_{23}\mathcal{H}(\bm q)\Gamma_{23}^{-1}=\mathcal{H}(-\bm q),\\
&\Gamma_{25}\hat{K}\mathcal{H}(\bm q)\left(\Gamma_{25}\hat{K}\right)^{-1}=\mathcal{H}(-\bm q)
\end{align}
where $\Gamma_{23}=-i\Gamma_2\Gamma_3$, $\Gamma_{25}=-i\Gamma_2\Gamma_5$, and $\hat{K}$ represents the complex conjugation. The joint anti-unitary symmetry $\mathcal{K}=\Gamma_{23}\Gamma_{25}\hat{K}=-\Gamma_{35}\hat{K}$ leads to
\begin{align}
\mathcal{K}\mathcal{H}(\bm q)\mathcal{K}^{-1}=\mathcal{H}(\bm q),
\end{align}
which shows that each band at $\bm q$ point is doubly degenerate.
Here we introduce the time-dependent phonon displacement of the lowest acoustical mode at $K$ point with $l_{\text{ph}}=1$ as $(u_x,u_y)=u_A(\cos{\tau},\sin{\tau})$ and $(v_x,v_y)=u_B(-\cos{\tau},\sin{\tau})$.
The OAM operator is given by $\hat{L}_z=-\hbar s_z\otimes\sigma_z=-\hbar\Gamma_{23}$, and the geometric OAM is evaluated by
\begin{align}
L_{z,n}(\bm q,\tau)=&~\hbar\omega\sum_{m\neq n}\frac{\braket{\psi_{n\bm q}(\tau)|\hat{L}_z|\psi_{m\bm q}(\tau)}\braket{\psi_{m\bm q}(\tau)|i\partial_{\tau}\psi_{n\bm q}(\tau)}}{E_{n\bm q}(\tau)-E_{m\bm q}(\tau)}+\text{c.c}\nonumber\\
=&~-\hbar\omega\sum_{m\neq n}\frac{\hbar\braket{\psi_{n\bm q}(\tau)|\hat{\Gamma}_{23}|\psi_{m\bm q}(\tau)}\braket{\psi_{m\bm q}(\tau)|i\partial_{\tau}\mathcal{H}^{(l_{\text{ph}}=1)}_{\text{eff}}(\bm q,\tau)|\psi_{n\bm q}(\tau)}}{\left[E_{n\bm q}(\tau)-E_{m\bm q}(\tau)\right]^2}+\text{c.c},
\end{align}
where $\partial_{\tau}\mathcal{H}^{(l_{\text{ph}}=1)}_{\text{eff}}(\bm q,\tau)=\lambda u_r\cos{\tau}\Gamma_4-\lambda u_r\sin{\tau}\Gamma_5$ with $u_r=u_B-u_A$. Here one can verify $\mathcal{K}\Gamma_{23}\mathcal{K}^{-1}=-\Gamma_{23}$ and $\mathcal{K}\Gamma_{4,5}\mathcal{K}^{-1}=\Gamma_{4,5}$, which lead to $L_{z,n}(\bm q,\tau)=0$.
On the other hand, in the case of phonons at $K$ point with $l_{\text{ph}}=0,-1$, the intervalley-scattering terms break such symmetry, resulting in nonzero OAM.

\section{OAM of monolayer transition metal dichalcogenides}
\subsection{Electronic OAM estimation}
In the main text, we obtain the OAM as a function of the wavenumber $q=\sqrt{q_x^2+q_y^2}$, which is given by
\begin{align}
\overline{L}_z(q)=\frac{-\tilde{\lambda}_u^4\hbar^2\omega}{(\tilde{v}^2q^2+\Delta^2+\tilde{\lambda}_u^2)^{\frac{3}{2}}(\tilde{v}^2q^2+\tilde{\lambda}_u^2)},
\end{align}
where $\Delta$ represents the energy gap.
The total OAM is given by the integration over the wavenumbers $q_x$ and $q_y$ as
\begin{align}
L^{\text{tot}}_z=\int_{-\infty}^{\infty}dq_x\int_{-\infty}^{\infty}dq_y\overline{L}_z(q_x,q_y)
\end{align}
by extending the region of the integration to infinity. Because the wavenumbers for $q_x$ and $q_y$ are isotropic, the integration can be calculated under the polar coordinate as
\begin{align}\label{Exact}
L^{\text{tot}}_z=\int_0^{2\pi}d\theta\int_0^{\infty}qdq\overline{L}_z(q)=-\frac{2\pi\tilde{\lambda}_u^4\hbar^2\omega}{\tilde{v}^2}\left[\frac{1}{\Delta^3}\text{arctanh}\left(\frac{\Delta}{\sqrt{\Delta^2+\tilde{\lambda}_u^2}}\right)-\frac{1}{\Delta^2\sqrt{\Delta^2+\tilde{\lambda}_u^2}}\right].
\end{align}
It shows that the OAM is inversely proportional to the cube of the energy gap.
Here $\tilde{\lambda}_u$ is a small quantity compared to the energy gap $|\Delta|$, and then the total OAM is approximately expressed as
\begin{align}\label{Lz_app}
L^{\text{tot}}_z\approx-\frac{2\pi\tilde{\lambda}_u^4\hbar^2\omega}{|\Delta|^3\tilde{v}^2}\text{ln}\left|\frac{2\Delta}{\tilde{\lambda}_u}\right|
\end{align}
with $\tilde{\lambda}_u=2f_1(u_{\text{X}}-u_{\text{M}})/a_0$ as mentioned in the main text. 

\begin{figure}
\begin{center}
\includegraphics[width=8cm]{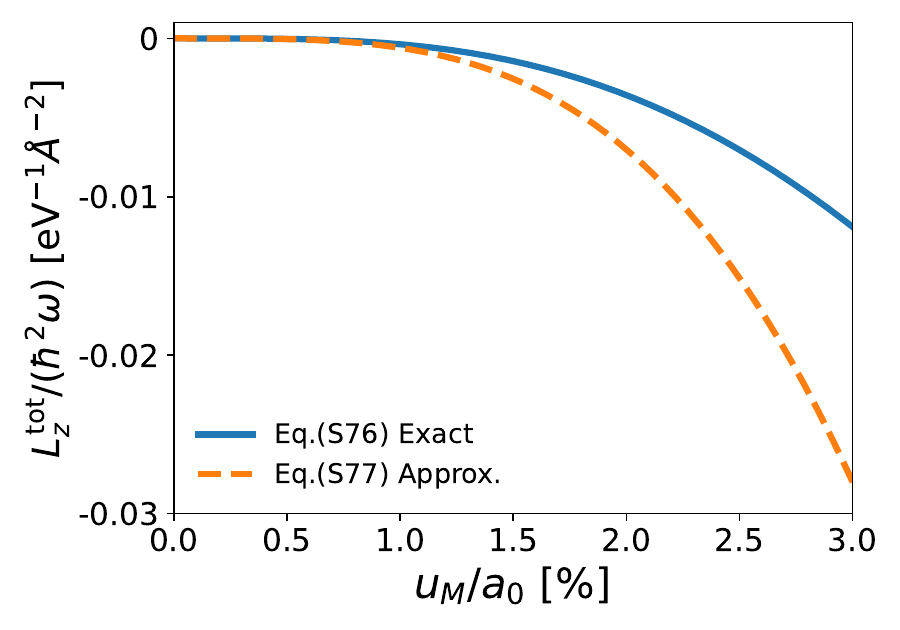}
\caption{Comparison of total OAM evaluated by Eqs.~(\ref{Exact}) and~(\ref{Lz_app}) for the TMD material WS$_2$.}
\label{sm_fig4}
\end{center}
\end{figure}

We compare the results of the total OAM evaluated by the exact formula in Eq.(\ref{Exact}) and the approximated formula in Eq.~({\ref{Lz_app}}) in Fig.~(\ref{sm_fig4}). We notice that they agree well if the phonon displacement is sufficiently small, and they deviate from each other when the phonon displacement gradually increases. Considering the validity range of Eq.~(\ref{Lz_app}), we assume that the phonon displacement amplitude of the atom M is 1\% of the lattice constant as $u_{\text{M}}=0.01a_0$.

\subsection{Effect of spin-orbital coupling}
\begin{figure}
\begin{center}
\includegraphics[width=14cm]{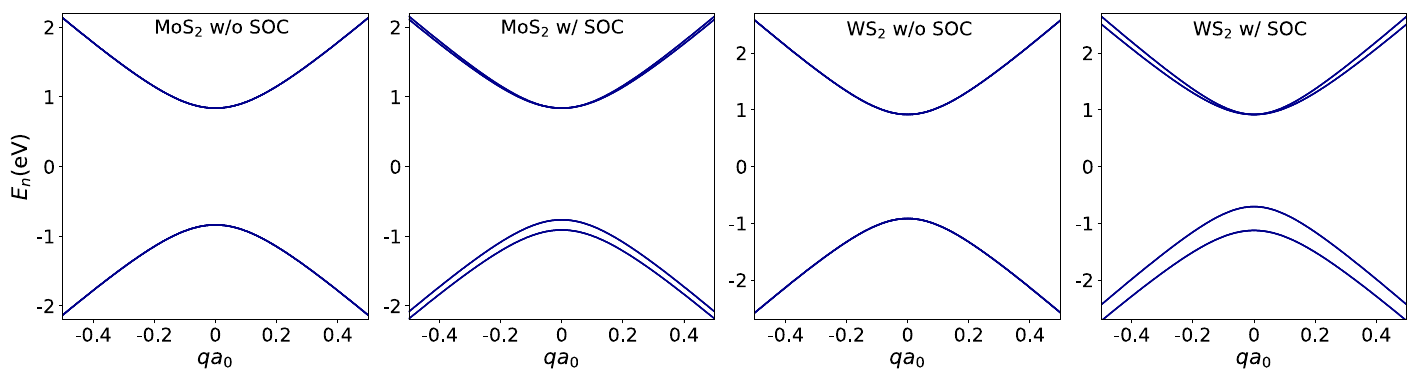}
\caption{Band structure of the effective Hamiltonian with and without SOC in 2D monolayer TMD materials MoS$_2$ and WS$_2$. Here the SOC strengthes $\lambda$ for MoS$_2$ and WS$_2$ are $0.073$eV and $0.211$eV, respectively~\cite{Liu2013}.}
\label{sm_fig5}
\end{center}
\end{figure}

Our proposed effect does not require spin-orbital coupling (SOC). Even though the SOC in 2D monolayer TMD materials MX$_2$ is substantial, the selection rule between phonons and electron orbitals is similar in the presence of SOC since the phonon-induced intervalley scattering does not flip the electron spins.
Here we consider the SOC only from the onsite contribution of the heavy transition-metal M atom. We extend our basis with the spin degree of freedom, which is $(\ket{K,d_{2},\uparrow},\ket{K,d_0,\uparrow},\ket{K',d_{-2},\uparrow},\ket{K',d_0,\uparrow},\ket{K,d_{2},\downarrow},\ket{K,d_0,\downarrow},\ket{K',d_{-2},\downarrow},\ket{K',d_0,\downarrow})$, the Hamiltonian of the SOC is given by
\begin{align}
\mathcal{H}'=\frac{\lambda}{\hbar^2}\bm L\cdot\bm S=\frac{\lambda}{2\hbar}s_z\otimes\hat{L}_z,
\end{align}
where $s_z$ denotes the Pauli matrix, $\hat{L}_z=\text{diag}(2\hbar,0,-2\hbar,0)$ represents the $z$ component of the OAM operator under the basis of $(\ket{K,d_{2}},\ket{K,d_0},\ket{K',d_{-2}},\ket{K',d_0})$, and $\lambda$ characterizes the SOC strength in the unit of eV. Note that the matrix elements of $\hat{L}_x$ and $\hat{L}_y$ are all zeros under the above basis.
Then the effective Hamiltonian after considering the SOC is written as
\begin{align}\label{Hsoc}
\mathcal{H}_{\text{soc}}=s_0\otimes\mathcal{H}_{\text{eff}}+\frac{\lambda}{2\hbar}s_z\otimes\hat{L}_z=\begin{bmatrix}
\mathcal{H}_{\text{eff}}+\frac{\lambda}{2\hbar}\hat{L}_z & \\
& \mathcal{H}_{\text{eff}}-\frac{\lambda}{2\hbar}\hat{L}_z
\end{bmatrix},
\end{align}
where $\mathcal{H}_{\text{eff}}$ is the effective Hamiltonian given in the main text. We show the band structures of the typical TMD materials MoS$_2$ and WS$_2$ with and without SOC in Fig.~\ref{sm_fig4}. In the absence of SOC, each band is fourfold degeneracy, which consists of the electronic Bloch states at $K$ and $K'$ points with spin degeneracy. The induced OAM should be the double from two spin channels. When the SOC is switched on, a small spin splitting comparing to the energy gap occurs as shown in Fig.~\ref{sm_fig5}.
We notice that even though the SOC is taken into account, the phonon-induced intervalley scattering of electron for up-spin and down-spin channels are independent from Eq.~(\ref{Hsoc}). Meanwhile, the induced OAM depends on the inverse cubic of energy gap $\Delta$ as shown in Eq.~(\ref{Lz_app}). Since the spin splitting $\lambda$ from the SOC is much smaller than the energy gap $\Delta$, the difference of the induced OAM between up-spin and down-spin channels can be negligible.

\section{OAM in strontium titanate}
A recent study has reported a transient orbital magnetization induced by chiral phonons in a molecular model of strontium titanate SrTiO$_3$~\cite{Urazhdin2025}, where the phonons are gradually switched on from $t=-\infty$ to $t=0$, and the electronic OAM is calculated by the time-dependent perturbation theory. Since the ways of introducing the phonons are different between Ref.~\cite{Urazhdin2025} and our paper, in this section, we directly calculate the OAM in such molecular model by using our approach to compare with Ref.~\cite{Urazhdin2025}.

\begin{figure}
\begin{center}
\includegraphics[width=12cm]{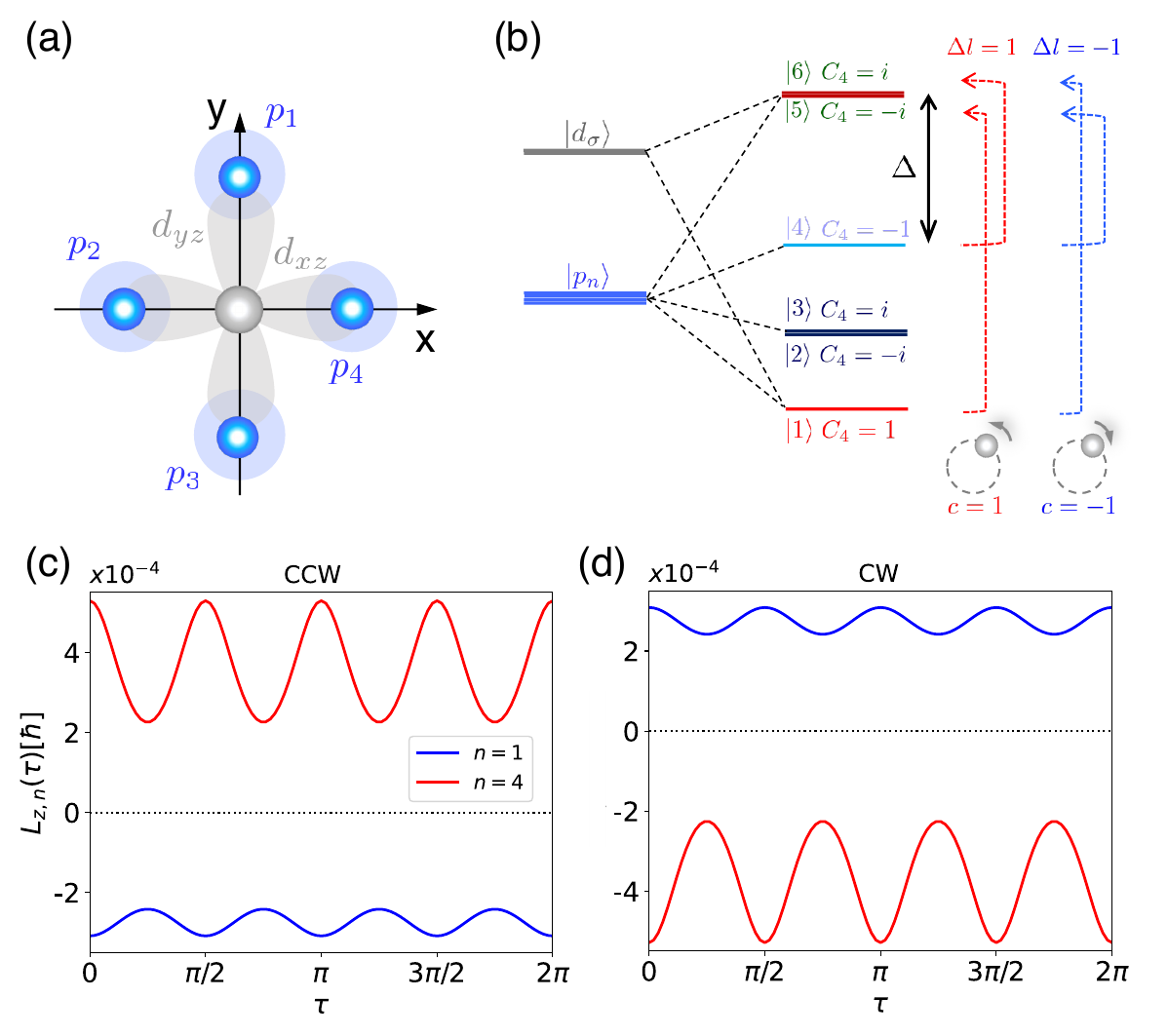}
\caption{Electronic OAM induced by chiral phonons in SrTiO$_3$. (a) Schematic of orbital configuration in the $xy$ plane, where the $d_{xz}$ and $d_{yz}$ orbitals of the central Ti atom hybridize with four surrounding oxygen $p_z$ orbitals. (b) Symmetry-resolved hybridization between the Ti $d_{\sigma}$ and oxygen $p_n$ orbitals classified by their $C_4$ eigenvalues, and the crystal-field splitting $\Delta$ separates the $C_4=\pm1$ and $C_4=\pm i$ sectors. Dashed lines indicate the phonon-induced transitions of electrons with CCW ($\Delta l=1$) and CW ($\Delta l=-1$) phonon modes. (c) and (d) Numerical results of the dynamical OAM during one period of phonon dynamics. Red and blue lines represent the contributions from the occupied states $\ket{1}$ and $\ket{4}$, respectively.}
\label{sm_fig6}
\end{center}
\end{figure}

The molecular model describing SrTiO$_3$ consists of the $p_z$ orbitals of the four oxygen atoms and the $d_{\sigma}=-(\sigma d_{xz}+id_{yz})/\sqrt{2}$ $(\sigma=\pm1)$ orbitals of Ti~\cite{Urazhdin2025} as shown in Fig.~\ref{sm_fig6}(a). We assume that the onsite energeis of these orbitals as $\bra{p_n}\hat{H}\ket{p_n}=0$ $(n=1,\cdots,4)$ and $\bra{d_{\sigma}}\hat{H}\ket{d_{\sigma}}=E_d$, and the hopping terms between O-O and Ti-O atoms are given by
\begin{align}
&\hat{H}_{\text{O-O}}=t_p\sum_{n=1}^3\left(\ket{p_n}\bra{p_{n+1}}+\ket{p_{n+1}}\bra{p_{n}}\right),\\
&\hat{H}_{\text{Ti-O}}=t\sum_{n=1}^4\sum_{\sigma=\pm1}\left(e^{i\frac{n\pi\sigma}{2}}\ket{p_{n}}\bra{d_{\sigma}}+e^{-i\frac{n\pi\sigma}{2}}\ket{d_{\sigma}}\bra{p_{n}}\right),
\end{align}
with their hopping parameters $t_p$ and $t$. Under the basis $(\ket{p_1},\ket{p_2},\ket{p_3},\ket{p_4},\ket{d_+},\ket{d_-})^{\mathsf{T}}$, the total Hamiltonian is written as
\begin{align}
\hat{H}_0=\begin{pmatrix}
0 & t_p & 0 & t_p & it & -it \\
t_p & 0 & t_p & 0 & -t & -t \\
0 & t_p & 0 & t_p & -it & it \\
t_p & 0 & t_p & 0 & t & t \\
-it & -t & it & t & E_d & 0 \\
it & -t & -it & t & 0 & E_d
\end{pmatrix}.
\end{align}
Here we can check the Hamiltonian preserves $C_4$ rotation symmetry as $\hat{C}_4\hat{H}_0\hat{C}_4^{-1}=\hat{H}_0$ by using $\hat{C}_4\ket{d_{\sigma}}=e^{i\frac{\pi\sigma}{2}}\ket{d_{\sigma}}$ and $\hat{C}_4\ket{p_n}=\ket{p_{n+1}}$.
We diagonalize the total Hamiltonian by employing the parameters as $t_p=-0.8$eV, $t=1.14$eV, and $E_d=3.7$eV~\cite{Urazhdin2025}, and the six eigenstates and their $C_4$ eigenvalues are given by
\begin{align}
&\ket{1}=\left(\frac{1}{2},\frac{1}{2},\frac{1}{2},\frac{1}{2},0,0\right)^\mathsf{T},~C_4=1\\
&\ket{2}=\left(\frac{i}{2}\cos{\theta_b},-\frac{1}{2}\cos{\theta_b},-\frac{i}{2}\cos{\theta_b},\frac{1}{2}\cos{\theta_b},\sin{\theta_b},0\right)^\mathsf{T},~C_4=-i\\
&\ket{3}=\left(-\frac{i}{2}\cos{\theta_b},-\frac{1}{2}\cos{\theta_b},\frac{i}{2}\cos{\theta_b},\frac{1}{2}\cos{\theta_b},0,\sin{\theta_b}\right)^\mathsf{T},~C_4=i\\
&\ket{4}=\left(-\frac{1}{2},\frac{1}{2},-\frac{1}{2},\frac{1}{2},0,0\right)^\mathsf{T},~C_4=-1\\
&\ket{5}=\left(\frac{i}{2}\cos{\theta_a},-\frac{1}{2}\cos{\theta_a},-\frac{i}{2}\cos{\theta_a},\frac{1}{2}\cos{\theta_a},\sin{\theta_a},0\right)^\mathsf{T},~C_4=-i\\
&\ket{6}=\left(-\frac{i}{2}\cos{\theta_a},-\frac{1}{2}\cos{\theta_a},\frac{i}{2}\cos{\theta_a},\frac{1}{2}\cos{\theta_a},0,\sin{\theta_a}\right)^\mathsf{T},~C_4=i\\
\end{align}
with their energies from low to high
\begin{align}
&E_1=2t_p=-1.6\text{eV},~(t_p<0)\\
&E_2=E_3=\frac{E_d}{2}-\sqrt{\frac{E_d^2}{4}+4t^2}=-1.09\text{eV},\\
&E_4=-2t_p=1.6\text{eV},\\
&E_5=E_6=\frac{E_d}{2}+\sqrt{\frac{E_d^2}{4}+4t^2}=4.79\text{eV},
\end{align}
where $\theta_{a,b}=\arctan{\left(\frac{2t}{-\frac{E_d}{2}\pm\sqrt{\frac{E_d^2}{4}+4t^2}}\right)}$.

When chiral phonons are present, the hopping parameters between Ti-O atoms are dynamically modulated. We assume that this modulation is characterized by a single displacement vector $\bm u$ of oxygen atoms relative to the Ti atom. The displacement vector $\bm u$ can be projected onto two perpendicular directions $\bm e_{l,n}=(\cos{\frac{n\pi}{2}},\sin{\frac{n\pi}{2}})$ and $\bm e_{t,n}=(-\sin{\frac{n\pi}{2}},\cos{\frac{n\pi}{2}})$ as
\begin{align}
u_{l,n}=\bm u\cdot\bm e_{l,n}=u\cos\xi_n,\\
u_{t,n}=\bm u\cdot\bm e_{t,n}=u\sin\xi_n,
\end{align}
where $\xi_n=\varphi_n-\frac{n\pi}{2}$ with $\varphi_n$ being the polar angle of instantaneous displacement~\cite{Urazhdin2025}. Here the presence of $u_{l,n}$ and $u_{t,n}$ modulates the hopping parameters between Ti-O atoms in the change of bond length: $\delta t^{(l)}_{n,\sigma}\propto u_le^{i\frac{n\pi\sigma}{2}}$ and also in the change of phase: $\delta t^{(t)}_{n,\sigma}\propto i\sigma u_te^{i\frac{n\pi\sigma}{2}}$, respectively. Thus, the total change of the hopping parameter is given by
\begin{align}
\delta t_{n,\sigma}=ue^{i\frac{n\pi\sigma}{2}}\left(a_l\cos\xi_n+i\sigma a_t\sin\xi_n\right),
\end{align}
where $a_l$ and $a_t$ are the proportionality constant along the $\bm e_l$ and $\bm e_n$ directions, respectively, with the dimension of force~\cite{Urazhdin2025}. It can be further calculated as
\begin{align}
\delta t_{n,\sigma}=ua_+e^{i\sigma\varphi}+(-1)^nua_-e^{-i\sigma\varphi},
\end{align}
where $a_{\pm}=(a_l\pm a_t)/2$, and $\varphi=c\omega t$ denotes the time dependence of circularly polarized phonons with $c=\pm 1$ corresponding to the CCW and CW modes. Therefore, we can write the modulated Hamiltonian as
\begin{align}
\delta \hat{H}(t)=\sum_{n=1}^{4}\sum_{\sigma=\pm 1}\left(\delta t^*_{n,\sigma}\ket{p_{n}}\bra{d_{\sigma}}+\delta t_{n,\sigma}\ket{d_{\sigma}}\bra{p_{n}}\right),
\end{align}
which preserves the $C_4$ rotation symmetry followed by $\frac{c\pi}{2\omega}$ time translation as
\begin{align}
\hat{C}_4\delta\hat{H}(t)\hat{C}^{-1}_4=\delta\hat{H}(t+\frac{c\pi}{2\omega}),
\end{align}
and $c$ also represents the phonon pseudo-angular momentum.
The modulated Hamiltonian in matrix representation is given by
\begin{align}
\delta \hat{H}(t)=\begin{pmatrix}
0 & 0 & 0 & 0 & \delta t^*_{1,+} & \delta t^*_{1,-} \\
0 & 0 & 0 & 0 & \delta t^*_{2,+} & \delta t^*_{2,-} \\
0 & 0 & 0 & 0 & \delta t^*_{3,+} & \delta t^*_{3,-} \\
0 & 0 & 0 & 0 & \delta t^*_{4,+} & \delta t^*_{4,-} \\
\delta t_{1,+} & \delta t_{2,+} & \delta t_{3,+} & \delta t_{4,+} & 0 & 0 \\
\delta t_{1,-} & \delta t_{2,-} & \delta t_{3,-} & \delta t_{4,-} & 0 & 0
\end{pmatrix}.
\end{align}

As is different from Ref.~\cite{Urazhdin2025}, here we consider that the phonon amplitude does not change with time. 
The OAM for the $n$th occupied states ($n=1,4$) is given by
\begin{align}\label{Lzunocc}
L_{z,n}(t)=\sum_{m=5,6}^{\text{unocc}}\left(\frac{\hbar\bra{n(t)}\hat{L}_z\ket{m(t)}\bra{m(t)}i\partial_t\ket{n(t)}}{E_n(t)-E_m(t)}+\text{c.c}\right),
\end{align}
where $\hat{L}_z=\text{diag}(0,0,0,0,-\hbar,\hbar)$ is the OAM operator, and $\ket{n(t)}$ is the $n$th eigenstates of the total Hamiltonian $\hat{H}(t)=\hat{H}_0+\delta\hat{H}(t)$.
We numerically calculate the induced OAM in Eq.~(\ref{Lzunocc}) as shown in Figs~\ref{sm_fig6}(c) and~\ref{sm_fig6}(d), where the time averages of the OAM from the occupied states $\ket{1}$ and $\ket{4}$ are around $-3c\times 10^{-4}\hbar$ and $4c\times 10^{-4}\hbar$, respectively.
This numerical result can be understood in terms of the virtual transitions which satisfy the selection rule between electrons and phonons, and one can see this in terms of the perturbation theory. These processes are shown in Fig.~\ref{sm_fig6}(b), where only the electrons in the occupied states $\ket{1}$, $\ket{4}$ can be excited to the unoccupied state $\ket{5}$, $\ket{6}$ because of the change of pseudo-angular momentum $\Delta l=c=\pm1$.

In order to compare with the results obtained from the time-dependent perturbation theory in Ref.~\cite{Urazhdin2025}, here we also calculate the induced OAM in Eq.~(\ref{Lzunocc}) by using the time-dependent perturbation theory. To describe the phonon-induced transition process, it suffices to retain the four states: $\ket{1}$, $\ket{4}$, $\ket{5}$, and $\ket{6}$, while the two states $\ket{2}$ and $\ket{3}$ are neglected. The phonon-perturbed energy states are then given by
\begin{align}
\text{occupied states}:&\ket{1'(t)}=\ket{1}+\frac{2a_+u\sin\theta_ae^{ic\omega t}}{4t_p-\Delta}\ket{5}+\frac{2a_+u\sin\theta_ae^{-ic\omega t}}{4t_p-\Delta}\ket{6},\\
&\ket{4'(t)}=\ket{4}+\frac{2a_-u\sin\theta_ae^{-ic\omega t}}{-\Delta}\ket{5}+\frac{2a_-u\sin\theta_ae^{ic\omega t}}{-\Delta}\ket{6},\\
\text{unoccupied states}:&\ket{5'(t)}=\ket{5}+\frac{2a_+u\sin\theta_ae^{-ic\omega t}}{\Delta-4t_p}\ket{1}+\frac{2a_-u\sin\theta_ae^{ic\omega t}}{\Delta}\ket{4},\\
&\ket{6'(t)}=\ket{6}+\frac{2a_+u\sin\theta_ae^{ic\omega t}}{\Delta-4t_p}\ket{1}+\frac{2a_-u\sin\theta_ae^{-ic\omega t}}{\Delta}\ket{4},
\end{align}
where $\Delta\equiv\frac{E_d}{2}+\sqrt{\frac{E_d^2}{4}+4t^2}+2t_p=3.2$eV is the energy gap as shown in Fig.~\ref{sm_fig6}(b). These perturbed states are proportional to the first order of the phonon amplitude $u$. By using Eq.~(\ref{Lzunocc}), the induced OAM in the occupied states $n=1,4$ are given by
\begin{align}\label{Lz1}
&L_{z,1}=-c\hbar\omega\frac{16a_+^2u^2\hbar\sin^4\theta_a}{(\Delta-4t_p)^3}\approx-5c\times10^{-4}\hbar,\\ \label{Lz4}
&L_{z,4}=c\hbar\omega\frac{16a_-^2u^2\hbar\sin^4\theta_a}{\Delta^3}\approx12c\times10^{-4}\hbar,
\end{align}
with the parameters $a_+=13$eV/nm and $a_-=7$eV/nm, the phonon amplitude $u=0.08$nm, and the phonon energy $\hbar\omega=12.4$meV~\cite{Urazhdin2025}. 
Comparing the results calculated from the time-dependent perturbation theory in Eqs.~(\ref{Lz1}) and (\ref{Lz4}) with those in Figs.~\ref{sm_fig6}(c) and ~\ref{sm_fig6}(d), we find that the numerical results are smaller than those calculated from perturbation theory but with the same order and sign.

We also compare these results of the OAM in Eqs.~(\ref{Lz1}) and (\ref{Lz4}) with those in the previous study~\cite{Urazhdin2025}, where they show a similar behavior with the previous study, such as the inverse cubic dependence on the energy gap $\Delta$, and the quadratic dependence on the phonon amplitude $u$, while we find that the resulting expression differ by a constant prefactor. The reason for this difference in the prefactor remains to be figured out, and it is left as a future work.


%